\pdfoutput=1
\documentclass{article} % For LaTeX2e
\usepackage{iclr2025_conference,times}

\usepackage{amsmath,amsfonts,bm}

\def\eqref#1{equation~\ref{#1}}
\def\1{\bm{1}}

\DeclareMathAlphabet{\mathsfit}{\encodingdefault}{\sfdefault}{m}{sl}
\SetMathAlphabet{\mathsfit}{bold}{\encodingdefault}{\sfdefault}{bx}{n}

\usepackage{hyperref}
\usepackage{url}
\usepackage{graphicx}
\usepackage{booktabs}
\usepackage{multirow}
\usepackage{array}
\usepackage{colortbl}
\usepackage{siunitx}
\usepackage{pifont}
\usepackage{xcolor}
\usepackage{enumitem}
\usepackage[normalem]{ulem}
\useunder{\uline}{\ul}{}

\newcommand{\specialcell}[2][c]{%
  \begin{tabular}[#1]{@{}c@{}}#2\end{tabular}}
\newcommand{\specialcellleft}[2][c]{%
  \begin{tabular}[#1]{@{}l@{}}#2\end{tabular}}

\usepackage{amssymb}
\usepackage{pifont}% http://ctan.org/pkg/pifont
\newcommand{\cmark}{\ding{51}}%
\newcommand{\xmark}{\ding{56}}% 

\usepackage[compact]{titlesec}
\usepackage{sidecap}
\titlespacing*{\section}{0pt}{*0.3}{*0.3}
\titlespacing*{\subsection}{0pt}{*0.3}{*0.3}
\titlespacing*{\subsubsection}{0pt}{*0.3}{*0.3}
\newcommand\AbMethods[1]{\textbf{#1}}
\newenvironment{xQuote}{%
  \par
  \setlength{\leftskip}{1.5em}%
  \setlength{\rightskip}{1.5em}%
}{%
  \par
}

\title{ProteinBench: A Holistic Evaluation of Protein Foundation Models}

\author{Fei Ye\thanks{Equal contribution.}~, Zaixiang Zheng$^*$, Dongyu Xue$^*$, Yuning Shen$^*$, Lihao Wang$^*$\\
\textbf{Yiming Ma, Yan Wang, Xinyou Wang, Xiangxin Zhou} and \textbf{Quanquan Gu}\thanks{Corresponding author.} \\
ByteDance Research \\
\texttt{\{yefei.joyce,quanquan.gu\}@bytedance.com} \\
\\Project page: \url{https://proteinbench.github.io/}
}

\iclrfinalcopy % Uncomment for camera-ready version, but NOT for submission.
\begin{document}
% \nolinenumbers

\maketitle

\begin{abstract}
Recent years have witnessed a surge in the development of protein foundation models, significantly improving performance in protein prediction and generative tasks ranging from 3D structure prediction and protein design to conformational dynamics. However, the capabilities and limitations associated with these models remain poorly understood due to the absence of a unified evaluation framework. To fill this gap, we introduce ProteinBench, a holistic evaluation framework designed to enhance the transparency of protein foundation models. Our approach consists of three key components: (i) A taxonomic classification of tasks that broadly encompass the main challenges in the protein domain, based on the relationships between different protein modalities; (ii) A multi-metric evaluation approach that assesses performance across four key dimensions: quality, novelty, diversity, and robustness; and (iii) In-depth analyses from various user objectives, providing a holistic view of model performance. Our comprehensive evaluation of protein foundation models reveals several key findings that shed light on their current capabilities and limitations. To promote transparency and facilitate further research, we release the evaluation dataset, code, and a public leaderboard publicly for further analysis and a general modular toolkit. We intend for ProteinBench to be a living benchmark for establishing a standardized, in-depth evaluation framework for protein foundation models, driving their development and application while fostering collaboration within the field. 
% The benchmark is available at \url{https://proteinbench.github.io/}.
\end{abstract}

\section{Introduction}
Proteins are fundamental molecules playing pivotal roles in a vast array of biological processes, from enzymatic catalysis and signal transduction to structural support and immune response. Their amino acid sequences determine their functions, often mediated through folding into specific three-dimensional structures. Understanding the complex interplay between protein sequence, structure, and function is crucial for advancing science and engineering spanning pharmaceuticals, agriculture, specialty chemicals, and biofuels~\citep{Kuhlman2019}.

In recent years, there has been a surge in the development of protein foundation models\footnote{In this study, we broaden the definition of protein foundation models to include any generative models aimed at addressing foundational problems of protein sciences.} aimed at understanding fundamental biological processes by capturing the intricate mechanisms of proteins~\citep{Jumper2021,abramson2024accurate,lin2023,watson2023rfdiffusion,ingraham2023illuminating,Krishna2024,Shin2021,Madani2023,Alley2019,wang2024diffusion,hayes2024simulating,Hie2024}. These models, leveraging advanced deep-learning and generative AI techniques, have demonstrated remarkable capabilities and marks a significant shift from traditional, task-specific approaches to more generalizable frameworks capable of learning complex patterns and relationships within vast protein datasets. For instance,  AlphaFold3~\citep{abramson2024accurate}, which is based on diffusion models, has achieved unprecedented accuracy in full atom structure prediction for all biomolecules, while others like the ESM series~\citep{Rives2021,Hsu2022,lin2023,verkuil2022language,hayes2024simulating} and DPLM \citep{wang2024diffusion} have shown impressive representation capability in protein language modeling benefiting diverse downstream tasks. Furthermore, these foundation models are not limited to single modalities. Multi-modal models that jointly consider sequence, structure, and function are emerging, offering a comprehensive understanding of protein behavior~\citep{hayes2024simulating,haiyan2023diffusion}. One important aspect of understanding this sequence-structure-function relationship is protein conformational dynamics. Recent work has extended protein structure prediction to several conformation prediction tasks and introduced generative AI to model the conformational distribution of proteins~\citep{jingEigenFoldGenerativeProtein2023,zheng2024predictingDiG,jingAlphaFoldMeetsFlow2023,wang2024proteinconfdiff,luStr2StrScorebasedFramework2024}. 

However, the rapid progress of protein foundation models has also led to an urgent need for a unified framework to holistically evaluate their performance across a diverse set of tasks, datasets, and metrics, as shown in Appendix \ref{appendix:sum_benchmarks}. The current landscape of protein foundation models is characterized by ununified modeling approaches, task-specific or model-specific evaluation criteria. This heterogeneity in evaluation methods makes it challenging to draw meaningful comparisons between different models and to fully understand their relative strengths and limitations.

Through systematic evaluations of datasets spanning diverse biological domains, with a particular emphasis on protein design and conformational dynamics, we aims to provide a comprehensive analysis of model architecture and performance on protein foundation models. This approach allows us to dissect the impact of various model components and data characteristics on different aspects of protein modeling. Comparing the capabilities of these models on standardized benchmarks is crucial for guiding future research directions, informing model selection for practical applications, and driving the advancement of the field as a
whole.

In this study, as shown in Figure \ref{fig:protbench}, we present ProteinBench, the first benchmark designed to provide a comprehensive evaluation of protein foundation models through four key components:

\begin{figure}
    \centering
    \includegraphics[width=\textwidth]{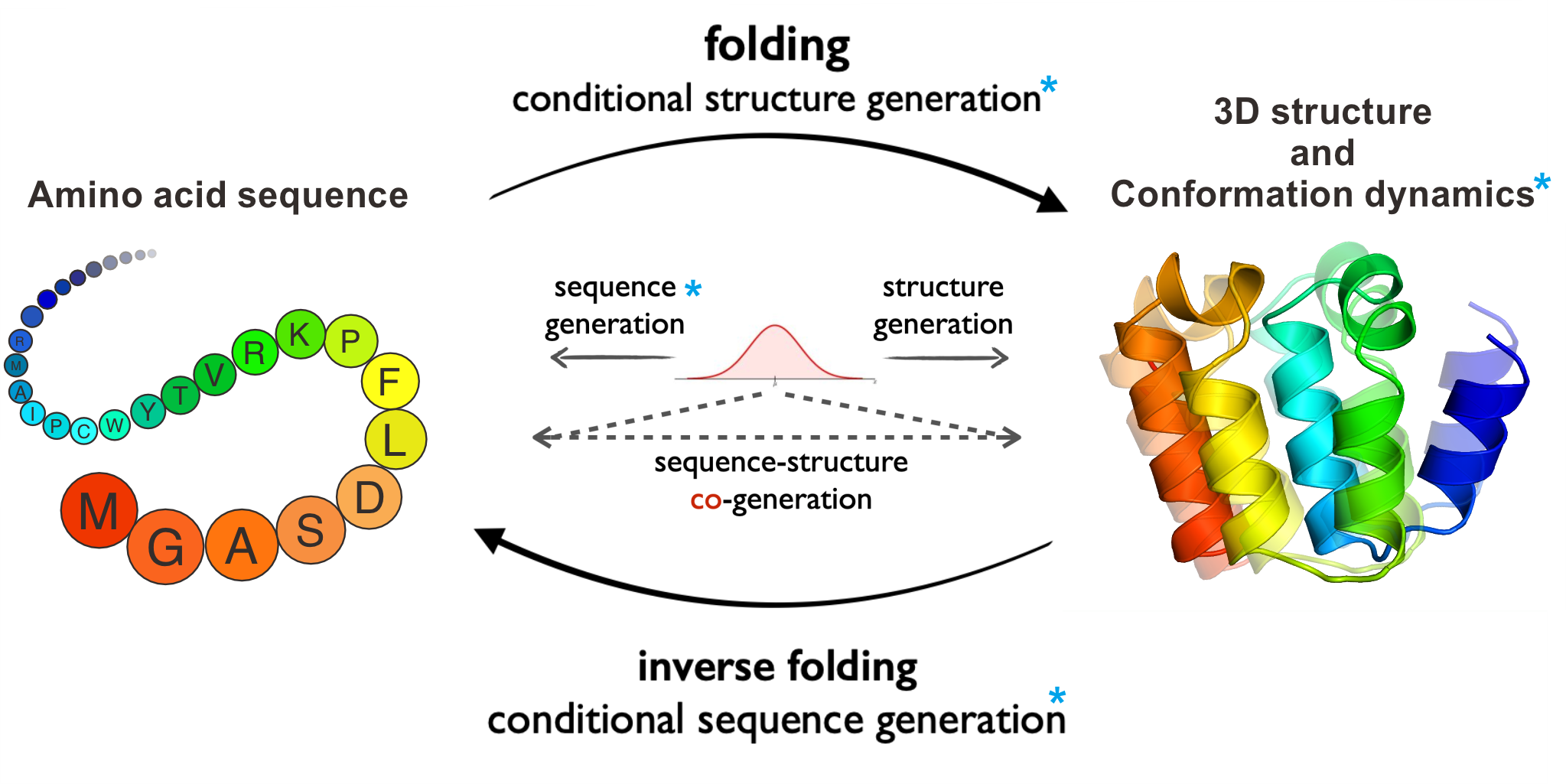}
    \caption{\textbf{Comprehensive overview of fundamental protein modeling tasks in ProteinBench.} ProteinBench incorporates a spectrum of protein modeling challenges. Tasks marked with blue stars highlight domains where standardized performance benchmarks were previously unavailable.} 
    \label{fig:protbench}
\end{figure}

\textbf{(1) A taxonomic classification of tasks encompassing the main generative challenges in the protein domain.} ProteinBench covers a wide range of generative tasks, including protein design (spanning structure design, sequence design, structure-sequence co-design, and an application-specific task of antibody design), three-dimensional structure prediction, and conformation dynamics. These tasks, addressing different protein modalities, enable a nuanced analysis of the interplay between model architecture and modal characteristics on performance. We utilize diverse and carefully curated datasets to capture the complexity and diversity of the protein universe, ensuring a thorough evaluation of model capabilities.

\textbf{(2) A multi-metric evaluation approach assessing performance across four key dimensions: quality, novelty, diversity, and robustness.} Current evaluations of protein generative models often suffer from non-unified metrics and incomplete assessments, typically focusing on only one or two aspects. However, protein scientific problems encompass a complex and systematic array of challenges. Downstream tasks in protein modeling and design involve intricate interplays between sequence, structure, and function. ProteinBench addresses this limitation by providing a comprehensive measurement of a model's ability to capture the mechanisms of the protein universe. We evaluate models based on four critical dimensions: quality, novelty, diversity, and robustness. This multi-faceted approach offers a more holistic view of model performance and capabilities.

\textbf{(3) In-depth analyses from various user objectives, providing a holistic view of model performance.}
Recognizing that different users may have varying objectives when applying protein foundation models, we conduct in-depth analyses from multiple perspectives. For instance, in protein design, some users may prioritize models that fit natural evolutionary distributions, while others may seek models capable of generating novel proteins outside the training set distribution. By analyzing model capabilities from these different objectives, ProteinBench provides insights that are beneficial for a wide range of practical applications.

\textbf{(4) Leaderboard and code framework.}
To facilitate fair comparisons and support the development of new methods, we provide a unified experimental framework. This includes a public leaderboard and open-source code, enabling researchers to easily benchmark their models against existing ones and contribute to the ongoing advancement of the field.

By incorporating these four components, ProteinBench aims to establish a standardized, comprehensive, and user-centric evaluation framework for protein foundation models. This approach not only illuminates the current state-of-the-art but also guides future research directions and accelerates progress in the field of protein modeling and design.

\section{Background and task definition}

\label{gen_inst}
In this section, we provide a concise overview of the tasks addressed by various protein foundation models as shown in Table \ref{tab:task_overview}, with a particular focus on two key generative tasks: protein design and conformational dynamics. These two areas are further divided into eight subtasks. 
\begin{table}[t]\footnotesize
    \centering
    \setlength{\tabcolsep}{2pt}
    \caption{Overview of ProteinBench, which summarizes the dimensions, metrics, and methods used in ProteinBench. We use ‘\textit{italics}’ for highlighting, a method that has not yet been evaluated in ProteinBench but will be assessed in the future.} 
    % The symbol '$^{*}$' highlights tasks that share the same designed structure evaluation pipeline, where the structures undergo inverse folding with ProteinMPNN and are then refolded using ESMFold. The maximum TM score to PDB and maximum cluster metrics are calculated using Foldseek. scTM is an abbreviation for self-consistent TM-score.}
    \resizebox{\textwidth}{!}{%
    \begin{tabular}{l|ll|l}
    \toprule
    \textbf{Tasks}  & \textbf{Dimension} & \textbf{Metrics} & \textbf{Methods}  \\
    \midrule
    \multicolumn{4}{l}{\textbf{\textit{Protein Design}}}\\
    \midrule
    \multirow{3}{*}{Inverse Folding}  & Sequence recovery & AAR & ProteinMPNN, ESMIF1, \\
    & Refoldability & scTM (AF2) & LM-Design, ESM3\\ 
    & Stability & pLDDT (AF2) & \textit{PiFold}, \textit{CarbonDesign}\\
    
    \cmidrule{2-4} \multirow{3}{*}{Backbone Design} & Quality  & scTM, scRMSD (ProteinMPNN \& ESMFold) & Rfdiffusion, Frameflow, Chroma,\\
    & Novelty & Max. TM score to PDB database (Foldseek) &   Framediff, Foldflow, Genie\\
    & Diversity & Pairwise TM, Max Cluster (Foldseek) & \textit{foldingdiff}, \textit{Proteus} \\

    \cmidrule{2-4} \multirow{3}{*}{Sequence Design} & Quality & pLDDT (AF2) & ProGen2, EvoDiff,  \\
    & Novelty & Max. TM to PDB database (Foldseek) & DPLM, ESM3 \\
    & Diversity & Pairwise TM , Max Cluster (Foldseek) & \\

    \cmidrule{2-4} \multirow{3}{*}{Struct-seq Co-design} & Quality  & scTM, scRMSD (ESMFold) & ProteinGenerator, ProtPardelle,  \\
    & Novelty & Max. TM score to PDB database (Foldseek) & Multiflow, ESM3, \textit{CarbonNovo} \\
    & Diversity & Pairwise TM, Max Cluster (Foldseek) & \\
    
    \cmidrule{2-4} \multirow{1}{*}{Motif Scaffolding} & Quality & Motif RMSD, Scafold RMSD  & FrameFlow, Rfdiffusion, TDS, EvoDiff, DPLM, ESM3 \\
    
    \cmidrule{2-4} \multirow{5}{*}{Antibody Design} & Accuracy  & AAR, RMSD, TM-score & HERN, \\
    & Functionality & Binding Energy (Rosetta) & MEAN, dyMEAN, \\ 
    & Specificity & Seq Similarity, PHR & DiffAb, AbDPO \\ 
    & Rationality & CN-Score, Clashes, Seq Naturalness & \\ 
    &  & Total Energy (Rosetta), scRMSD (IgFold) & \\ 
    \midrule
    \multicolumn{4}{l}{\textbf{\textit{Protein Conformation Prediction}}}\\ 
    \midrule
    \multirow{2}{*}{\specialcellleft[c]{Single state\\(folding)}}  
        & Accuracy & TM score, RMSD, GDT, lDDT  & AlphaFold2, OpenFold, ESMFold, \\
        & Quality  & CA clash/break rate, Peptide bond break rate        &  RosettaFold2, EigenFold  \\
    \cmidrule{2-4}
    \multirow{3}{*}{\specialcellleft[c]{Multiple state\\Prediction}}  
        & Accuracy  & Ensemble TM score/RMSD  & \multirow{6}{*}{\specialcellleft[c]{EigenFold, MSA-subsampling,\\Str2Str, AlphaFlow/ESMFlow,\\ConfDiff}} \\
        & Diversity & pairwise RMSD/TM & \\ 
        & Quality   & CA clash/break rate, Peptide bond break rate                 & \\ 
    \cmidrule{2-3}
    \multirow{3}{*}{\specialcellleft[c]{Distribution\\Prediction}}         
        & Accuracy  & Flexibility accuracy, Distributional similarity, Ensemble observables & \\
        & Diversity & Pairwise RMSD, RMSF                                                   & \\
        & Quality   & CA clash/break rate, Peptide bond break rate                               & \\
    \bottomrule
    \end{tabular}%
    }%
    \label{tab:task_overview}%
\end{table}
For each task, we focus on the following aspects, with detailed information provided in the appendix:
\begin{xQuote}
\textbf{[Task Definition]} A clear and concise description of the task, including its objectives and relevance to protein science. Specification of the input data format and expected output for each task. \\
\textbf{[Evaluation Metrics]} Description of the metrics used to assess model performance, including quality, novelty, diversity, and robustness measures. \\
\textbf{[Datasets]} Overview of the datasets used for each task, including their size, diversity, and any pre-processing steps applied. 
\end{xQuote}

\subsection{Protein Design}

\subsubsection{Inverse folding}
%\begin{xQuote}
\textbf{[Task Definition]} The objective is to predict an optimal amino acid sequence for a given target protein structure, considering factors such as stability, refoldability, and potential functionality. \\
\textbf{[Evaluation Metrics]} Performance in protein sequence design is assessed using multiple complementary metrics: \textbf{(1) Sequence Recovery}: This metric compares the designed sequences to natural sequences with similar structures. It quantifies how well the design method can recapitulate evolutionarily conserved sequence patterns associated with specific structural motifs. \textbf{(2) Refoldability}: This measure evaluates the structural similarity between the target backbone and the predicted structure of the designed sequence. The prediction is performed using AlphaFold2~\citep{Jumper2021}. Similarity is quantified using self-consistent template modeling score (scTM)~\citep{trippe2022diffusion} and self-consistent root-mean-square deviation (scRMSD), providing insight into how well the designed sequence would fold into the intended structure. \textbf{(3) Stability}: This is assessed using the predicted local distance difference test (pLDDT) calculated by AlphaFold2. The pLDDT score serves as a proxy for the predicted stability of the designed protein, which is used in ~\cite{Dauparas2022}. \\
\textbf{[Datasets]} Evaluations were conducted on different datasets targeting two distinct objectives of structure-based sequence design: \textbf{(1) capture the native evolutionary distribution}: we evaluated two independent datasets containing newly released PDB structures: CASP15~\citep{casp15} and CAMEO~\citep{Robin2021}. We collected new structures from the ongoing CAMEO assessment between January and July 2024, resulting in a total of 332 complex structures. Additionally, 32 protein structures were collected from CASP15, which includes only protein entities, excluding nucleic acids or ligands. \textbf{(2) de novo protein design}: RFdiffusion~\citep{Watson2023} was used to generate backbones of varying lengths: specifically, 100, 200, 300, 400, and 500 residues. For each length, 10 different structures were randomly sampled, using a sampling temperature of 0.1 for all methods. The designability of these sequences was evaluated using AlphaFold2, with the scTM score and pLDDT metrics serving as the primary assessment criteria. Existing benchmarks for inverse folding, such as PDB-Struct~\citep{wang2023pdb} and Proteininvbench~\citep{gao2024proteininvbench}, provide standardized protein structure sets for evaluating inverse folding methods. While these benchmarks have significantly contributed to the field's advancement, there is a growing need for more comprehensive evaluation frameworks. These expanded evaluations should align more closely with diverse user objectives in protein design, encompassing aspects like accuracy in capturing natural evolutionary distributions and robustness in de novo backbone-based sequence design. 
%\end{xQuote}

\subsubsection{Protein backbone design}
%\begin{xQuote}
\textbf{[Task Definition]} Protein backbone design focuses on creating new protein folds to achieve de novo design objectives. This task is essential for expanding the repertoire of protein structures beyond those found in nature, with significant applications in fields such as drug discovery, biomaterials, and therapeutics. \\
\textbf{[Evaluation Metrics]} The evaluation of backbone design encompasses multiple criteria to assess both the quality and novelty of generated structures. Structural \textbf{quality} is primarily measured using self-consistent TM-score and RMSD, which provide quantitative measures of the backbone's refoldability measured by ProteinMPNN~\citep{Dauparas2022} and ESMFold~\citep{lin2023}. Equally important are \textbf{novelty} metrics, which gauge the method's capacity to explore new structural space beyond known protein folds. This aspect is evaluated using two key metrics: The maximum TM-score obtained when comparing designed structures to existing entries in the RCSB Protein Data Bank (PDB)~\citep{berman2000rcsb}. This comparison is performed using Foldseek~\citep{van2022foldseek}, a rapid structural alignment tool. \textbf{Diversity} metrics, which include: (a) Pairwise maximum TM-scores among the designed structures. (b) The number of distinct structural clusters identified within the set of designed backbones, also determined using Foldseek~\citep{van2022foldseek}. These diversity metrics help quantify the range of unique structures the design method can produce, ensuring that it's not simply recreating known folds but generating a varied repertoire of protein backbones. \\
\textbf{[Datasets]} The primary objective of generative tasks is to accurately map the general distribution of the training set. For protein structure generation, high-resolution structures from the Protein Data Bank (PDB) are commonly used. To gain insight into this data distribution, we randomly sampled 100 native single-chain structures from the RCSB database as references. To ensure diversity, we iteratively removed structures with the highest TM-score compared to others, until we arrived at a final set of 100 distinct structures. This approach provides a representative snapshot of the single-chain structural distribution within the PDB, serving as a benchmark for evaluating the performance of generative models in capturing the true distribution of protein structures.
%\end{xQuote}

\subsubsection{Protein sequence design}
%\begin{xQuote}
\textbf{[Task Definition]} The aim of this task is to generate amino acid sequences of desired properties, such as quality, diversity and novelty. 
Besides sequence-based evaluation, the structural characteristics of the generated sequences are also important.   \\
\textbf{[Evaluation Metrics]} For sequence naturalness, we use perplexity from an autoregressive protein language model (ProGen2) to quantify if the patterns of generated sequences lie in natural sequence distribution. 
% We use sequence clustering based on MMseq2 to measure the inner-similarity among generated samples (diversity) of and similarity to the dataset (novelty). 
For structure-based evaluation, we use single-sequence folding model, i.e., ESMFold, to predict the structure of the generated sequences, and then measure the structural quality by pLDDT as the proxy of \textbf{structural stability} of the sequence using their predicted structures from AlphaFold2, as well as structural diversity and novelty using the same protocol as in backbone design. 
\\
\textbf{[Datasets]} UniRef50 is the commonly used dataset for training protein sequence generative models and language models.  
%\end{xQuote}

\subsubsection{Structure and sequence co-design}
%\begin{xQuote}
\textbf{[Task Definition]} Protein structure-sequence co-design involves simultaneously optimizing both the backbone structure and amino acid sequence of a protein to achieve desired properties or functions. This task is more complex than sequence design or structure design alone, as it explores a larger solution space. \\
\textbf{[Evaluation Metrics]} Evaluation metrics are derived from those used for both sequence and structure design: structure quality assessments, sequence-structure compatibility, as well as novelty of both sequence and structure compared to known proteins is also crucial. \\
\textbf{[Datasets]} High-resolution protein structures from the Protein Data Bank (PDB) is the commonly used datasets for this task, with careful consideration given to remove redundancy. 
%\end{xQuote}

\subsubsection{Motif scaffolding}
\textbf{[Task Definition]} Motif scaffolding involves designing a protein structure that incorporates a specific functional motif or binding site. The goal is to create a stable protein framework (scaffold) that presents the desired motif in the correct geometry for its function. \\
\textbf{[Evaluation Metrics]} Following \citet{yim2024improved}, key metrics include the structural accuracy of the motif within the designed scaffold (typically measured by RMSD), overall protein stability, and retention of the motif's functional properties. Experimental validation through binding assays or enzymatic activity tests is often crucial. \\
\textbf{[Datasets]} Datasets typically include libraries of known functional motifs (e.g., catalytic sites, binding interfaces) and diverse scaffold structures that can potentially accommodate these motifs. 
The Protein Data Bank is a primary source, but curated datasets of functional sites like the Catalytic Site Atlas are also valuable. \\
\textbf{[Related benchmarks]} Enzyme Design Challenge provides relevant test cases. However, given the specificity of motif scaffolding tasks, benchmarks often need to be tailored to the particular class of motifs or functions being targeted. Currently, there exists no comprehensive benchmark for this task in the field. 
A widely used benchmark containing 17 (25) motif-scaffolding problems was used in RFDiffusion~\citep{watson2023rfdiffusion}.

\subsubsection{Antibody Design}
\textbf{[Task Definition]} The goal of antibody design is to generate antibodies that can specifically bind to a given antigen. Since the Complementarity-Determining Regions (CDRs) of antibodies are highly variable and primarily responsible for antigen binding, antibody design could be simplified to the design of CDR regions and further reduced to the design of the third CDR in heavy chain (CDR-H3). Given the crucial role that protein structure plays in interactions, antibody design usually involves the simultaneous design of the sequence and the structure when binding to the antigen. \\
\textbf{[Evaluation Metrics]} As a highly goal-oriented functional protein design task, the evaluation of antibody design is straightforward, namely the \textbf{Functionality} (binding capability to the target antigen) and \textbf{Specificity} of the designed antibody. Additionally, the \textbf{Rationality} of the designed antibodies sequence and structure needs to be evaluated for filtering out invalid designs. Existing studies also evaluate the \textbf{Accuracy} of designed antibodies by measuring their similarity to natural antibodies as natural ones are confirmed to be effective. However, using accuracy as an evaluation metric is inadequate in many cases, which we will demonstrate in detail in Section \ref{sec:antibody_evaluation}. \\
\textbf{[Datasets]} The Structural Antibody Database (SAbDab \cite{10.1093/nar/gkt1043}) is the commonly used dataset for antibody design. It contains structural data of the antibody-antigen complex, but the data size is limited and contains numerous redundancies.

\subsection{Protein Conformation Prediction} 

\subsubsection{Protein Folding: single-state prediction}
\textbf{[Task Definition]} Protein folding is the task of predicting the folded structure of a protein from its sequence. Folding models, such as AlphaFold2, have played a pivotal role in the recent development of models for protein conformation prediction \citep{jingAlphaFoldMeetsFlow2023,wang2024proteinconfdiff}. Therefore, we recognize the necessity of including protein folding in this benchmark, viewing it as a specific instance of protein conformation prediction for a single conformational state.\\
\textbf{[Evaluation Metrics]} The \textbf{accuracy} of a predicted structure is evaluated by compared with its reference structures deposited in PDB using RMSD, TM-score, global distance test (GDT), and local distance difference test (lDDT). We also evaluate the \textbf{quality} of predicted structures by measuring the rate of clashing alpha carbons (CA-clash), disconnecting neighbor alpha carbons (CA-break), and disconnecting peptide bonds (PepBond-break) in predicted structures. See Appendix \ref{apdx:conf_metrics} for details. \\
\textbf{[Datasets]} Most of the folding models compared in this benchmark were established prior to 2022. We use CAMEO2022 from \cite{jingEigenFoldGenerativeProtein2023} for evaluation, which consists of 183 short-to-mid-length single protein chains ($<$ 750 amino acids) from the targets of CAMEO between Aug 1 and Oct 31, 2022. % Periodic or continuing prediction assessment such as CASP and CAMEO. Recent models such as AlphaFold3 chooses to collect "recent test dataset" from the records deposited after a set cutoff date for evaluations. 

\subsubsection{Multiple-state prediction}
\textbf{[Task Definition]} As an extension of the single-state prediction task, multiple-state prediction aims to accurately predict (by sampling) two or more distinct conformational states of a protein that have been observed under different conditions (e.g., ligand binding) or through molecular dynamics simulations. The ability to predict these ``alternative'' conformations in addition to the folded structure could provide insights into conformational changes and protein functions.
\\
\textbf{[Evaluation Metrics]} We evaluate this task based on accuracy, diversity, and quality. The \textbf{accuracy} of predicting a state is determined by the best structural similarity of the samples to the reference structure, measured by TM-score or RMSD. The overall accuracy of multiple-state prediction is assessed by ``ensemble accuracy'', which is the average accuracy across all reference states (TMens or RMSDens where ``ens'' stands for ensemble), similar to \cite{jingEigenFoldGenerativeProtein2023}. For sample structural \textbf{diversity}, we measure the pairwise TM-score (or RMSD) among the samples. Finally, we assess the structural \textbf{quality} of generated samples, similar to single-state prediction, using CA-clash, CA-break, and PepBond-break.\\
\textbf{[Datasets]} 
We benchmark the models on two public datasets from previous works: 1) \textit{apo-holo}, which contains 91 proteins, each with a pair of experimental structures (\textit{apo} or unbound, and \textit{holo} or bound) related to ligand-binding-induced conformational changes \citep{saldano2022impactapo,jingEigenFoldGenerativeProtein2023}; (2) BPTI (Bovine Pancreatic Trypsin Inhibitor), a 58 amino acids protein, where a previous long-time MD simulation revealed five clusters of distinct conformations \citep{shaw2010atomic}.
% \textbf{[Related benchmarks]} While recent methods are sporadically compared in their papers, there lacks a unified and comprehensive benchmark of these methods encompasing a variety of tasks.

\subsubsection{Distribution prediction}
\textbf{[Task Definition]} In contrast to multiple-state prediction, where the main goal is to recover specific conformational states, distribution prediction focuses on generating a sample distribution that resembles a target distribution—such as the empirical distribution sampled from molecular dynamics (MD). This task further bridges the gap between protein conformation prediction models and current MD-based approaches for studying protein dynamics and thermodynamic properties.\\
\textbf{[Evaluation Metrics]} In addition to the \textbf{quality} and \textbf{diversity} criteria from the previous sections, we follow \citep{jingAlphaFoldMeetsFlow2023} and include three categories of metrics to compare the ensemble of model-generated samples with the reference samples from MD simulations: \textbf{flexibility} assesses whether the model can distinguish more ``flexible'' regions or proteins from less ``flexible'' ones, measured by the Pearson correlation $r$ of region/protein diversity (e.g., pairwise RMSD); \textbf{distributional accuracy} directly compares the conformational distributions of model-generated samples with the reference MD conformations through the Wasserstein distance or the cosine similarity of the first principal components; and \textbf{ensemble observables} focus on function-related observables, such as transient contacts between residues due to dynamics, and compare the sample ensemble with the reference ensemble from MD. See Appendix \ref{apdx:conf_metrics} for detailed descriptions of the metrics.\\
\textbf{[Datasets]} We evaluate performance using the ATLAS dataset \citep{yannATLASProteinFlexibility2024}, a recent database of MD simulation results for diverse proteins. To avoid data leakage for models trained on portions of the ATLAS dataset, we follow \cite{jingAlphaFoldMeetsFlow2023} and benchmark on 82 proteins whose PDB entries were deposited after May 1, 2019 and are not part of the training or validation set.
% \textbf{[Related benchmarks]} While recent methods are sporadically compared in their papers, there lacks a unified and comprehensive benchmark of these methods encompasing a variety of tasks.

\section{ProteinBench}
In this section, we provide ProteinBench, a holistic evaluation framework for protein foundation models. By systematically evaluating protein foundation models on the following tasks, we aim to provide a comprehensive understanding of their capabilities and limitations. This approach allows for a nuanced comparison of different model architectures and strategies, highlighting areas of strength and identifying opportunities for improvement. All data used in this benchmark are publicly available, ensuring reproducibility and facilitating wider participation in the research community. 

\subsection{Protein Design}
In this section, we present a comprehensive evaluation of various protein foundation models across fundamental protein design tasks, including single-modal approaches (structure-based sequence design, structure design, and sequence design), multi-modal structure-sequence co-design, and the application-specific task of antibody design. This holistic assessment allows us to examine the versatility and effectiveness of different modeling approaches across a spectrum of protein engineering challenges. Notably, for backbone design, sequence design, co-design, and motif scaffolding, the quality, novelty, and diversity metrics are calculated using the same method. By utilizing common evaluation metrics across tasks, we enable cross-task comparisons, hoping to provide performance analysis to identify the strengths and limitations of each modeling approaches, and help to uncover potential synergies between different protein modals for future research. 

\subsubsection{Inverse folding}
In this section, we evaluate the performance of various inverse-folding models for structure-based sequence design, focusing on two distinct objectives: natural evolutionary fitness (in-distribution proteins) and de novo designed backbone-based sequence design. The latter represents an out-of-distribution problem that tests the robustness of the methods, as these structures typically contain some noise different from high-resolution structure deposited in PDBs. The results are presented in Table \ref{tab:inverse_folding}.

Our analysis of native distribution fitness reveals that language model-based methods, for example \textsc{LM-Design}~\citep{zheng2023structure} in our investigation, effectively capture the natural evolutionary distribution, achieving high sequence recovery rates for native protein structure-based sequence design. This suggests that these models effectively learn and replicate the intricate patterns of amino acid selection that have emerged through evolutionary processes. However, its performance decreases when applied to de novo backbone-based sequence design. Conversely, ProteinMPNN~\citep{Dauparas2022}, a method specifically developed for de novo design and trained using coordinates perturbed with 0.2Å added noise, consistently demonstrates superior performance in de novo design tasks. However, ProteinMPNN's performance shows a decline when evaluated on objective to fiting to native evolution. This finding has significant implications for the field, suggesting that no single model currently excels across all protein design objectives. The choice of model should be carefully aligned with the intended applications. 

ESM-IF1~\citep{Hsu2022} was trained on the largest dataset of native sequences and structures from the AlphaFoldDB~\citep{varadi2022alphafold} based on GVP~\citep{jing2020gvp} and Transformer architectures and incorporated 0.1Å noise during training (similar to ProteinMPNN), it showed suboptimal performance in de novo backbone sequence design. Further investigation into the effects of larger noise additions or alternative model architectures on ESM-IF1's performance could prove insightful. It is worth noting that we did not include functional mutation prediction tasks in this study, an area where ESM-IF1 has demonstrated impressive results, as these have been extensively studied in other benchmarks, such as ProteinGYM~\citep{notin2024proteingym}. ESM3~\citep{hayes2024simulating}, the recently released multi-modal protein language model, exhibits performance comparable to ESM-IF1, with notable advantages for specific sequence lengths (100, 300, and 400 residues). We have noticed that certain inverse folding methods such as PiFold~\citep{gao2022pifold} and CarbonDesign~\citep{ren2024accurate} are currently not featured in ProteinBench. We plan to update their performance soon in the near future.
% Table generated by Excel2LaTeX from sheet 'BCR'
\begin{table}[t]\footnotesize
  \centering\setlength{\tabcolsep}{2pt}
  \caption{Performance of structure-based sequence design models on inverse folding tasks. The reported results are the median of repetitive experiments. ’N/A’ stands for not applicable. ESMIF1 and ESM3 use all native structures and sequences for model training, therefore, they not measured in the evolution distribution fitting objective.}
    \resizebox{\textwidth}{!}{%
    \begin{tabular}{lcc|cc|cc|cc|cc|cc}
    \toprule
      & \multicolumn{2}{c}{\textbf{Fitting Evolution Distribution}} & \multicolumn{10}{c}{\textbf{De novo backbones based sequence design }} \\

    \cmidrule{2-13}  & \multicolumn{1}{c}{\textbf{CASP}} & \multicolumn{1}{c}{\textbf{CAMEO}}  & \multicolumn{2}{c}{\textbf{length 100}} & \multicolumn{2}{c}{\textbf{length 200}} & \multicolumn{2}{c}{\textbf{length 300}} & \multicolumn{2}{c}{\textbf{length 400}} & \multicolumn{2}{c}{\textbf{length 500}} \\

    \cmidrule{2-13}  & \multicolumn{1}{c}{\textbf{AAR ↑}} & \multicolumn{1}{c}{\textbf{AAR ↑}}  & \multicolumn{1}{c}{\textbf{scTM ↑}} & \multicolumn{1}{c}{\textbf{pLDDT ↑}} & \multicolumn{1}{c}{\textbf{scTM ↑}} & \multicolumn{1}{c}{\textbf{pLDDT ↑}} & \multicolumn{1}{c}{\textbf{scTM ↑}} & \multicolumn{1}{c}{\textbf{pLDDT ↑}} & \multicolumn{1}{c}{\textbf{scTM ↑}} & \multicolumn{1}{c}{\textbf{pLDDT ↑}} & \multicolumn{1}{c}{\textbf{scTM ↑}} & \multicolumn{1}{c}{\textbf{pLDDT ↑}} \\
    \midrule

    ProteinMPNN & 0.450 & 0.468 & \textbf{0.962} & \textbf{94.14} & \textbf{0.945} & \textbf{89.34} & \textbf{0.962} & \textbf{90.28} & \textbf{0.875} & \textbf{83.76} & \textbf{0.568} & \textbf{67.09} \\
    ESM-IF1 & N/A & N/A & 0.810 & 88.83 & 0.635 & 69.67 & 0.336 & 74.36 & 0.449 & 64.59 & 0.462 & 58.97 \\
    \textsc{LM-Design} & \textbf{0.516} & \textbf{0.570} & 0.834 & 78.45 & 0.373 & 58.41 & 0.481 & 69.86 & 0.565 & 59.87 &0.397 & 56.35 \\
    % ByProt(ESM2 3b) & 0.565 & 0.610 & 0.357 & 76.00 & 0.416 & 58.09 & 0.332 & 58.72 & 0.337 & 57.85 & 0.290 & 51.92 \\
    ESM3 & N/A & N/A & 0.942 & 86.60 & 0.486 & 60.69 & 0.632 & 70.78 & 0.564 & 62.63 & 0.452 & 59.37 \\

    % MSA-1b & 0.921{\footnotesize$\pm$0.001} & 0.857$\pm$0.004 & 0.689$\pm$0.014 & 0.887$\pm$0.009 & 0.679$\pm$0.019 & 0.557$\pm$0.025 & 0.416 $\pm$ 0.050\\
    
    % \midrule

    \bottomrule
    \end{tabular}%
    }%
  \label{tab:inverse_folding}%
\end{table}%

\subsubsection{Structure design}
In this section, we evaluate the performance of protein foundation models for backbone design. The results are presented in Table \ref{tab:backbone_design}. Our analysis focuses on the quality, novelty, and diversity of the generated structures across various chain lengths. Based on the quality metrics of scTM-score and scRMSD, RFdiffusion~\citep{watson2023rfdiffusion} demonstrates exceptional performance in backbone design for chain lengths ranging from 50 to 300 amino acids. FrameFlow~\citep{yim2023fast} achieves the second-best performance in this range. However, we observe a significant performance decrease across all models for longer chains (500 amino acids), with scTM scores dropping by more than 20\%. This decline suggests that developing methods for long-chain backbone design remains an important challenge for future research. Novelty is an equally important metric, as it gauges a method's capacity to explore new structural space beyond known protein folds. Under moderate quality constraints (scTM score \textgreater 0.5), FoldFlow~\citep{bose2023se} and Genie~\citep{lin2023generating} exhibit good performance in generating novel structures. When we increase the quality threshold (scTM score \textgreater 0.8), Chroma~\citep{ingraham2023illuminating} generally shows the best performance across chain lengths from 50 to 500 amino acids. In terms of structural diversity, Chroma shows commendable performance across the tested chain lengths. It is important to note that for this evaluation, we used the released FoldFlow model trained on a smaller training set with shorter sequences. This limitation may lead to an unfair comparison of the model architecture to other methods trained on the entire PDB database, particularly for longer chain lengths. We will soon update our evaluations to include more methods, such as Foldingdiff~\citep{wu2024protein} and Proteous~\citep{wang2024proteus}.

% Table generated by Excel2LaTeX 
\begin{table}[t]\footnotesize
  \centering\setlength{\tabcolsep}{2pt}
  \caption{Performance of backbone design models evaluated using various lengths ranging from 50 to 500. The reported results are the median of repetitive experiments. We highlight the \textbf{best} performance in bold and the \uline{second-best} with the underline. For the novelty and diversity metrics, we only highlight results with the corresponding scTM score higher than 0.5. `N/A' stands for not applicable.}
    \resizebox{\textwidth}{!}{%
    \begin{tabular}{lccccc|ccccc}
    \toprule
      & \multicolumn{5}{c}{\textbf{length 50}} & \multicolumn{5}{c}{\textbf{length 100}}  \\

    \cmidrule(lr){2-6} \cmidrule(lr){7-11}   & \multicolumn{2}{c}{\textbf{Quality}} & \multicolumn{1}{c}{\textbf{Novelty}}  & \multicolumn{2}{c}{\textbf{Diversity}}  & \multicolumn{2}{c}{\textbf{Quality}} & \multicolumn{1}{c}{\textbf{Novelty}}  & \multicolumn{2}{c}{\textbf{Diversity}} \\

    \cmidrule(lr){2-6} \cmidrule(lr){7-11}  & \multicolumn{1}{c}{\textbf{scTM ↑}} & \multicolumn{1}{c}{\textbf{scRMSD ↓}}  & \multicolumn{1}{c}{\textbf{Max TM ↓}} & \multicolumn{1}{c}{\textbf{pairwise TM ↓}} & \multicolumn{1}{c}{\textbf{Max Clust. ↑}}  & \multicolumn{1}{c}{\textbf{scTM ↑}} & \multicolumn{1}{c}{\textbf{scRMSD ↓}}  & \multicolumn{1}{c}{\textbf{Max TM ↓}} & \multicolumn{1}{c}{\textbf{pairwise TM ↓}} & \multicolumn{1}{c}{\textbf{Max Clust.↑}} \\
    \midrule
    \rowcolor{gray!15} 
    Native PDBs  &      0.91$\pm$0.11  &        0.74$\pm$1.45 &         N/A      &         0.29$\pm$0.03 &          0.66 &         0.96$\pm$0.10  &        0.67$\pm$1.61       & N/A          & 0.30$\pm$0.02         &   0.77  \\
    RFdiffusion  &    \textbf{0.95$\pm$0.12}  &     \textbf{0.45$\pm$1.71} &         0.65$\pm$0.16 &         0.58$\pm$0.05 &          0.67 &        \textbf{0.98$\pm$0.12}  & \textbf{0.48$\pm$0.56} &    0.76$\pm$0.01       & 0.41$\pm$0.03            &   0.32  \\
    FrameFlow    &     \underline{0.91$\pm$0.09}  &  \underline{0.58$\pm$0.51} &         0.75$\pm$0.01 &         0.68$\pm$0.10 &          0.39 &    \underline{0.94$\pm$0.08}  &  \underline{0.70$\pm$0.70}       & 0.72$\pm$0.01     & 0.55$\pm$0.08         &   0.49  \\
    Chroma       &      0.85$\pm$0.15  &        1.05$\pm$1.49 &  \underline{0.59$\pm$0.08} &         0.29$\pm$0.01 &          0.48 &         0.89$\pm$0.13  &        1.27$\pm$1.85       & 0.70$\pm$0.01     & \underline{0.35$\pm$0.03}         &   0.59  \\
    FrameDiff(latest)    &      0.85$\pm$0.13  &        1.00$\pm$1.27 &         0.67$\pm$0.01 &         0.35$\pm$0.02 &          0.64 &         0.90$\pm$0.08  &        1.23$\pm$1.02       & 0.71$\pm$0.08     & 0.52$\pm$0.05         &   0.11  \\
    FoldFlow1(sfm)  &     0.90$\pm$0.10 &        0.67$\pm$0.88 &         0.68$\pm$0.03 &         0.63$\pm$0.07 &          0.48 &         0.87$\pm$0.11  &        1.34$\pm$1.42       & 0.65$\pm$0.01     & 0.49$\pm$0.08         &   \underline{0.83}  \\
    FoldFlow1(base) &     0.79$\pm$0.14 &        1.19$\pm$1.27 &         0.66$\pm$0.02 &    \underline{0.53$\pm$0.08} &          0.76 &         0.81$\pm$0.15  &        1.70$\pm$1.95       & \underline{0.62$\pm$0.01}     & 0.48$\pm$0.07         &  \underline{0.83}  \\
    FoldFlow1(ot)   &     0.83$\pm$0.16 &        1.10$\pm$1.53 &         0.65$\pm$0.02 &    \underline{0.53$\pm$0.08} &    \underline{0.77} &         0.83$\pm$0.15  &        1.60$\pm$1.95       & 0.64$\pm$0.01     & 0.48$\pm$0.06         &   0.81  \\
    Genie          &     0.57$\pm$0.15 &        3.12$\pm$2.07 &         \textbf{0.57$\pm$0.03} &  \textbf{0.32$\pm$0.02} &   \textbf{0.90} &         0.69$\pm$0.17  &        3.38$\pm$3.04    & \textbf{0.59$\pm$0.01}     & \textbf{0.31$\pm$0.02}        &   \textbf{0.96}  \\

    \midrule

      & \multicolumn{5}{c}{\textbf{length 300}} & \multicolumn{5}{c}{\textbf{length 500}}  \\

    \cmidrule(lr){2-6} \cmidrule(lr){7-11}  & \multicolumn{2}{c}{\textbf{Quality}} & \multicolumn{1}{c}{\textbf{Novelty}}  & \multicolumn{2}{c}{\textbf{Diversity}}  & \multicolumn{2}{c}{\textbf{Quality}} & \multicolumn{1}{c}{\textbf{Novelty}}  & \multicolumn{2}{c}{\textbf{Diversity}} \\

    \cmidrule(lr){2-6} \cmidrule(lr){7-11}  & \multicolumn{1}{c}{\textbf{scTM ↑}} & \multicolumn{1}{c}{\textbf{scRMSD ↓}}  & \multicolumn{1}{c}{\textbf{Max TM ↓}} & \multicolumn{1}{c}{\textbf{pairwise TM ↓}} & \multicolumn{1}{c}{\textbf{Max Clust. ↑}}  & \multicolumn{1}{c}{\textbf{scTM ↑}} & \multicolumn{1}{c}{\textbf{scRMSD ↓}}  & \multicolumn{1}{c}{\textbf{Max TM ↓}} & \multicolumn{1}{c}{\textbf{pairwise TM ↓}} & \multicolumn{1}{c}{\textbf{Max Clust.↑}} \\
    \midrule

    \rowcolor{gray!15} 
    Native PDBs  &      0.97$\pm$0.10 &         0.82$\pm$2.67 &         N/A        &  0.28$\pm$0.02 &   0.77  &  0.97$\pm$0.17 &   1.07$\pm$5.96 &  N/A       &   0.29$\pm$0.03   & 0.8   \\
    RFdiffusion  &      \textbf{0.96$\pm$0.15} &  \textbf{1.03$\pm$3.14} &         \underline{0.64$\pm$0.01}   &  \textbf{0.36$\pm$0.03} &   0.65  &  \textbf{0.79$\pm$0.19} &  \textbf{5.60$\pm$5.66} &  0.62$\pm$0.004 &   \underline{0.33$\pm$0.02}   & \underline{0.89}   \\
    FrameFlow    &     \underline{0.92$\pm$0.15} & \underline{1.95$\pm$2.76} &         0.65$\pm$0.01   &  0.43$\pm$0.07 &   \underline{0.88}  &  0.61$\pm$0.19 &   7.92$\pm$4.08 &  0.61$\pm$0.01 &   0.40$\pm$0.06    & 0.92   \\
    Chroma       &      0.87$\pm$0.13 &         2.47$\pm$3.63 &         0.66$\pm$0.01   &  \textbf{0.36$\pm$0.04} &   0.67  &  \underline{0.72$\pm$0.18} &   \underline{6.71$\pm$5.76} &  \underline{0.60$\pm$0.01} &   \textbf{0.29$\pm$0.01}    & \textbf{0.99}   \\
    FrameDiff(latest)    & 0.87$\pm$0.12 &         2.73$\pm$2.69 &         0.69$\pm$0.00  &  0.48$\pm$0.04 &   0.21  &  0.63$\pm$0.24 &   9.52$\pm$18.19&  \textbf{0.58$\pm$0.03}  &   0.40$\pm$0.06    & 0.52    \\
    FoldFlow1(sfm)  &      0.45$\pm$0.11 &         9.04$\pm$2.52 &       0.54$\pm$0.01   &  0.39$\pm$0.04 &   1.00  &  0.37$\pm$0.06 &   13.04$\pm$1.71&  0.53$\pm$0.01  &   0.37$\pm$0.03    & 1.00 \\
    FoldFlow1(base) &      0.43$\pm$0.09 &         9.56$\pm$2.42 &       0.54$\pm$0.01   &  0.39$\pm$0.05 &  0.98  &  0.35$\pm$0.05 &   13.20$\pm$2.29&  0.52$\pm$0.01  &   0.39$\pm$0.05    & 1.00  \\
    FoldFlow1(ot)   &      0.54$\pm$0.12 &         8.21$\pm$2.38 &       \textbf{0.58$\pm$0.00}  &  0.41$\pm$0.06 &   \textbf{0.94}  & 0.37$\pm$0.06 &    12.48$\pm$2.00&  0.51$\pm$0.01  &   0.35$\pm$0.03    & 1.00  \\
    Genie          &      0.27$\pm$0.02 &         20.37$\pm$1.70&       0.30$\pm$0.01   &  0.23$\pm$0.01 &   1.00  & 0.25$\pm$0.01 &    26.08$\pm$1.58&  0.22$\pm$0.002 &   0.23$\pm$0.004  & 1.00   \\
    % \midrule

    % MSA-1b & 0.921{\footnotesize$\pm$0.001} & 0.857$\pm$0.004 & 0.689$\pm$0.014 & 0.887$\pm$0.009 & 0.679$\pm$0.019 & 0.557$\pm$0.025 & 0.416 $\pm$ 0.050\\
    
    % \midrule

    \bottomrule
    \end{tabular}%
    }%
  \label{tab:backbone_design}%
\end{table}%

\subsubsection{Sequence design}
\begin{table}[t]\footnotesize
  \centering\setlength{\tabcolsep}{2pt}
  \caption{Performance of protein sequence generative models/language models on sequence generation tasks. The reported results are the average of repetitive experiments with the standard derivation. The pLDDT score is the output of AlphaFold2. Max TM is an abbreviation for Maximum TM-score to PDB database. 'N/A' stands for not applicable.We highlight the \textbf{best} performance in bold.}
\resizebox{\textwidth}{!}{%
\begin{tabular}{lccccc|ccccc}
\toprule
 % \multirow{2}{*}{Model}  
                         & \multicolumn{5}{c}{\textbf{length 100}}                               & \multicolumn{5}{c}{\textbf{length 200}}                                             \\
 \cmidrule(lr){2-6} \cmidrule(lr){7-11}
                             & \multicolumn{2}{c}{\textbf{Quality}}             & \multicolumn{2}{c}{\textbf{Diversity}}  & \multicolumn{1}{c}{\textbf{Novelty}}          & \multicolumn{2}{c}{\textbf{Quality} }            & \multicolumn{2}{c}{\textbf{Diversity}} & \multicolumn{1}{c}{\textbf{Novelty}}          \\
% \midrule
 \cmidrule(lr){2-6} \cmidrule(lr){7-11}

                       & \multicolumn{1}{c}{\textbf{ppl ↓}} & \multicolumn{1}{c}{\textbf{pLDDT ↑}}  & \multicolumn{1}{c}{\textbf{pairwise TM ↓}}  & \multicolumn{1}{c}{\textbf{Max Clust. ↑}} & \multicolumn{1}{c}{\textbf{Max TM ↓}} &  \multicolumn{1}{c}{\textbf{ppl ↓}}   & \multicolumn{1}{c}{\textbf{pLDDT↑}} & \multicolumn{1}{c}{\textbf{pairwise TM ↓}} & \multicolumn{1}{c}{\textbf{Max Clust. ↑}} & \multicolumn{1}{c}{\textbf{Max TM ↓}} \\
\midrule
\rowcolor{gray!15} 
Native Seqs  & & 68.46$\pm$16.50     & 0.55$\pm$0.19   & 0.75          & N/A        &          & 61.91$\pm$11.62     & 0.49$\pm$0.10 & 0.78         &  N/A                  \\
Progen 2 (700M)            &8.28$\pm$3.87 & 64.00$\pm$21.26   & \textbf{0.42$\pm$0.10} & 0.94         & \textbf{0.64$\pm$0.08}    &  5.68$\pm$3.64 & 69.91$\pm$9.23      & 0.40$\pm$0.13 & 0.91         & \textbf{0.69$\pm$0.05}       \\
EvoDiff                    &16.89$\pm$1.04 & 50.20$\pm$10.27     & 0.43$\pm$0.05  & \textbf{0.98}         & 0.69$\pm$0.03    &17.28$\pm$1.64   & 50.66$\pm$16.38     & 0.36$\pm$0.04 & 1.00             & 0.71$\pm$0.02       \\
DPLM (650M)         & \textbf{6.21$\pm$3.10}      & \textbf{85.38$\pm$14.20}     & 0.50$\pm$0.20  & 0.80      & 0.74$\pm$0.10   &\textbf{4.61$\pm$2.63 }        & \textbf{93.54$\pm$3.73}      & 0.54$\pm$0.24 & 0.70           & 0.91$\pm$0.004       \\
% DPLM2                        & 96.1735$\pm$1.2920      & 0.8918$\pm$0.2256  & 0.2           & 0.9451$\pm$0.0238       & 95.5912$\pm$1.0578      & 0.7054$\pm$0.2953 & 0.55          & 0.9682$\pm$0.0004       \\
% DPLM2\_update                & 94.1689$\pm$4.0755      & 0.5794$\pm$0.2590  & 0.75          &                     & 93.7375$\pm$5.8759      & 0.6351$\pm$0.2688 & 0.625         &                     \\
ESM3 (1.4B)                 & 14.79$\pm$2.90& 54.26$\pm$15.35     & 0.45$\pm$0.15  & 0.90           & 0.68$\pm$0.07       &12.96$\pm$2.38& 58.45$\pm$9.40      & \textbf{0.35$\pm$0.07} & \textbf{1.00}             & 0.80$\pm$0.01       \\
\midrule
                         & \multicolumn{5}{c}{\textbf{length 300}}                               & \multicolumn{5}{c}{\textbf{length 500}}                                             \\
 \cmidrule(lr){2-6} \cmidrule(lr){7-11}
                             & \multicolumn{2}{c}{\textbf{Quality}}             & \multicolumn{2}{c}{\textbf{Diversity}}  & \multicolumn{1}{c}{\textbf{Novelty}}          & \multicolumn{2}{c}{\textbf{Quality} }            & \multicolumn{2}{c}{\textbf{Diversity}} & \multicolumn{1}{c}{\textbf{Novelty}}          \\
% \midrule
 \cmidrule(lr){2-6} \cmidrule(lr){7-11}

                       & \multicolumn{1}{c}{\textbf{ppl ↓}} & \multicolumn{1}{c}{\textbf{pLDDT ↑}}  & \multicolumn{1}{c}{\textbf{pairwise TM ↓}}  & \multicolumn{1}{c}{\textbf{Max Clust. ↑}} & \multicolumn{1}{c}{\textbf{Max TM ↓}} &  \multicolumn{1}{c}{\textbf{ppl ↓}}   & \multicolumn{1}{c}{\textbf{pLDDT↑}} & \multicolumn{1}{c}{\textbf{pairwise TM ↓}} & \multicolumn{1}{c}{\textbf{Max Clust. ↑}} & \multicolumn{1}{c}{\textbf{Max TM ↓}} \\
\midrule
\rowcolor{gray!15} 
Native Seqs & & 61.49$\pm$14.47     & 0.51$\pm$0.13  & 0.85          & N/A            &   & 62.95$\pm$12.60     & 0.51$\pm$0.11 & 0.78         & N/A                  \\
Progen 2 (700M)        &6.25$\pm$ 4.02     & 65.69$\pm$20.93     & 0.42$\pm$0.16  & 0.93         & \textbf{0.66$\pm$0.06}   &4.27$\pm$3.60     & 61.45$\pm$20.17     & 0.32$\pm$0.11 & 0.95          & 0.68$\pm$0.08       \\
EvoDiff                    &17.13$\pm$2.00  & 45.14$\pm$9.95      & \textbf{0.31$\pm$0.03}  & \textbf{1.00}             & 0.68$\pm$0.02        & 16.51$\pm$3.82 &43.14$\pm$5.16     & \textbf{0.31$\pm$0.03} & \textbf{1.00}             & 0.69$\pm$0.02       \\
DPLM (650M)                &\textbf{3.47$\pm$1.44}  & \textbf{93.07$\pm$5.77}      & 0.57$\pm$0.25  & 0.63         & 0.91$\pm$0.01     &\textbf{3.33$\pm$1.8}  & \textbf{87.73$\pm$11.61}     & 0.43$\pm$0.18 & 0.85          & 0.85$\pm$0.04       \\
% DPLM2                        & 94.8445$\pm$1.2984      & 0.7154$\pm$0.2756  & 0.525         & 0.9517$\pm$0.0005       & 94.4432$\pm$1.6871      & 0.7847$\pm$0.2813 & 0.35          & 0.9650$\pm$0.0008       \\
% DPLM2\_update                & 94.9259$\pm$1.3688      & 0.5931$\pm$0.2541  & 0.65          &                     & 93.3041$\pm$3.5421      & 0.6158$\pm$0.2768 & 0.675         &                     \\
ESM3 (1.4B)                       &14.59$\pm$2.97 & 48.08$\pm$13.34     & 0.32$\pm$0.03  & \textbf{1.00}             & 0.75$\pm$0.02     & 11.10$\pm$2.26  & 52.17$\pm$10.52     & 0.30$\pm$0.05 & \textbf{1.00}             & \textbf{0.54$\pm$0.03}      \\
\bottomrule
\end{tabular}%
}%
\label{tab:seqdesign_struct}%
\end{table}%

In this section, we assess the performance of various protein sequence generative models based on the quality, diversity, and novelty of their generated sequences across different chain lengths. The evaluation metrics include AlphaFold2 (AF2) predicted pLDDT scores for structural plausibility (quality), maximum TM-score and maximum cluster values for structural diversity, and maximum TM-score to PDB structures for structural novelty.
We choose representative methods of distinct modeling foundations for evaluation.
Among the methods evaluated, ProGen2~\citep{Nijkamp2023} is an autoregressive protein language model (AR-LM), while EvoDiff~\citep{alamdari2023protein} is designed as an order-agnostic autoregressive diffusion model (OADM). 
DPLM~\citep{wang2024diffusion} and ESM3~\citep{hayes2024simulating} share a probabilistic foundation as absorbing discrete diffusion models or generative masked language models. 
Notably, ESM3 is a multimodal model that advances beyond other sequence-only methods by jointly learning protein sequences, structures, and functions through tokenization. 
For each model and sequence length, we sample 50 sequences to evaluate their performance.

As shown in Table~\ref{tab:seqdesign_struct}, DPLM consistently shows the highest quality scores, indicating superior accuracy in sequence generation. However, it has relatively lower diversity metrics, suggesting less variation in its generated sequences. EvoDiff, while demonstrating lower pLDDT scores, excels in diversity, particularly in producing highly diverse sequence clusters. Surprisingly, ESM3, a multimodal protein LM, displays lower pLDDT in sequence generation, while maintaining competitive diversity, especially in generating novel sequences. ProGen2 strikes a balance between quality and diversity, offering moderate pLDDT scores and satisfactory diversity and novelty. This model is effective for generating sequences that are both diverse and close to known structures, depending on specific application needs. Regarding different chain lengths, all the models generally exhibit consistent trends in their performance metrics. As the chain length increases, there is a slight decline in the quality of sequences generated by some models, particularly for EvoDiff and ESM3. This indicates a challenge in maintaining high sequence quality as the chain length grows. Among them, DPLM demonstrate robust performance across all lengths, maintaining high pLDDT even for longer sequences. Overall, DPLM is good at highly structural protein sequence generation, while EvoDiff and ESM3 are preferable for better diversity and novelty, with ProGen2 offering a balanced performance across metrics.

\subsubsection{Structure and sequence co-design}

\begin{table}[t]\footnotesize
\vspace{-1mm}
  \centering\setlength{\tabcolsep}{3pt}
  \caption{Performance of protein co-design models on structure-sequence co-generation tasks. The reported results are the average of repetitive experiments with the standard derivation. We highlight the best
performance in bold. \textbf{{$^*$}}: We have tried our best to reproduce all models according to the instructions in their respective codebases, using publicly available model weights. However, some results may differ from those reported in the original studies. We welcome any feedback and corrections to help us make timely updates in the future. }
\resizebox{\textwidth}{!}{%
\begin{tabular}{lcccc|cccc}
\toprule
                  & \multicolumn{4}{c}{\textbf{length 100}}                                     & \multicolumn{4}{c}{\textbf{length 200}}                                     \\
 \cmidrule(lr){2-5} \cmidrule(lr){6-9}
                  & \multicolumn{2}{c}{\textbf{Quality}} & \multicolumn{1}{c}{\textbf{Diversity}}     & \multicolumn{1}{c}{\textbf{Novelty}}           & \multicolumn{2}{c}{\textbf{Quality}} & \multicolumn{1}{c}{\textbf{Diversity}}     & \multicolumn{1}{c}{\textbf{Novelty}}            \\
                   \cmidrule(lr){2-5} \cmidrule(lr){6-9} & \multicolumn{1}{c}{\textbf{scTM ↑}}         & \multicolumn{1}{c}{\textbf{scRMSD ↓}}       & \multicolumn{1}{c}{\textbf{Max Clust. ↑}} & \multicolumn{1}{c}{\textbf{Max TM ↓}} & \multicolumn{1}{c}{\textbf{scTM ↑}}         & \multicolumn{1}{c}{\textbf{scRMSD ↓}}       & \multicolumn{1}{c}{\textbf{Max Clust. ↑}} & \multicolumn{1}{c}{\textbf{Max TM ↓}} \\
\midrule
\rowcolor{gray!15} 
Native PDBs & 0.91$\pm$0.11    & 2.98$\pm$3.49    & 0.75          & N/A                & 0.88$\pm$0.09    & 3.24$\pm$3.77    & 0.77          & N/A                \\
ProteinGenerator & 0.91$\pm$0.08    & 3.75$\pm$3.39    & 0.24          & 0.73                & 0.88$\pm$0.09    & 6.24$\pm$4.10    & 0.25          & 0.72                \\
ProtPardelle     & 0.91$\pm$0.09    & 2.07$\pm$1.87    & 0.73          & 0.16                & 0.92$\pm$0.04    & 2.36$\pm$1.19   & 0.09           & 0.75                \\
Multiflow         & \textbf{0.96$\pm$0.04}    & \textbf{1.10$\pm$0.71}    & 0.33          & 0.71                & \textbf{0.95$\pm$0.04}    & \textbf{1.61$\pm$1.73}    & 0.42          & 0.71                \\
ESM3*              & 0.72$\pm$0.19       & 13.80$\pm$10.51         & \textbf{0.64}          & \textbf{0.41}                & 0.63$\pm$0.20         & 21.18$\pm$16.19        & \textbf{0.63}          & \textbf{0.61}                \\
\midrule
             & \multicolumn{4}{c}{\textbf{length 300}}                                     & \multicolumn{4}{c}{\textbf{length 500}}                                     \\
 \cmidrule(lr){2-5} \cmidrule(lr){6-9}
                  & \multicolumn{2}{c}{\textbf{Quality}} & \multicolumn{1}{c}{\textbf{Diversity}}     & \multicolumn{1}{c}{\textbf{Novelty}}           & \multicolumn{2}{c}{\textbf{Quality}} & \multicolumn{1}{c}{\textbf{Diversity}}     & \multicolumn{1}{c}{\textbf{Novelty}}            \\
                   \cmidrule(lr){2-5} \cmidrule(lr){6-9} & \multicolumn{1}{c}{\textbf{scTM ↑}}         & \multicolumn{1}{c}{\textbf{scRMSD ↓}}       & \multicolumn{1}{c}{\textbf{Max Clust. ↑}} & \multicolumn{1}{c}{\textbf{Max TM ↓}} & \multicolumn{1}{c}{\textbf{scTM ↑}}         & \multicolumn{1}{c}{\textbf{scRMSD ↓}}       & \multicolumn{1}{c}{\textbf{Max Clust. ↑}} & \multicolumn{1}{c}{\textbf{Max TM ↓}} \\
\midrule
\rowcolor{gray!15} 
Native PDBs & 0.92$\pm$0.12    & 3.94$\pm$4.95    & 0.75          & N/A                & 0.90$\pm$0.14    & 9.64$\pm$7.05    & 0.80         & N/A                \\
ProteinGenerator & 0.81$\pm$0.14    & 9.26$\pm$4.13    & 0.22          & \textbf{0.71}                & 0.41$\pm$0.19    & 33.91$\pm$15.10    & 0.18          & 0.73                    \\
ProtPardelle     & 0.94$\pm$0.02    & 2.07$\pm$0.73   & 0.05          & 0.73                & 0.41$\pm$0.10    & 41.10$\pm$8.85   & 0.14           &  0.65                    \\
Multiflow         & \textbf{0.96$\pm$0.06}    & \textbf{2.14$\pm$3.24}    & \textbf{0.58}          & \textbf{0.71}                & \textbf{0.83$\pm$0.15}    & \textbf{8.48$\pm$7.02}    & \textbf{0.67}          & \textbf{0.68}                \\
ESM3*              & 0.59$\pm$0.21         & 25.5$\pm$20.68         & 0.52          & 0.73                & 0.54$\pm$0.20         & 33.70$\pm$21.08       & 0.37          & 0.77                \\
\bottomrule
\end{tabular}
}%
\label{tab:co-design}%
\end{table}

In this section, we examine the performance of protein structure-sequence co-generation, a topic that has recently gained significant interest within the research community.
We inspect the performance of ProteinGenerator~\citep{lisanza2023joint}, ProtPardelle~\citep{chu2024all}, Multiflow~\citep{campbellgenerative}, and ESM3~\citep{hayes2024simulating} for different lengths.
% [TODO: summarize the key properties of these models].
The performance is assessed using metrics similar to those applied in backbone generation.
Note that, however, the quality here is about structure-sequence compatibility measuring how well the designed sequence can fold into the corresponding designed structure, using scTM and scRMSD.
The key difference is that co-design models are tasked with simultaneously generating both the sequence and structure, while backbone design models require an additional inverse folding model, such as ProteinMPNN, to design the sequence. Other metrics used for evaluation include diversity (max cluster) and novelty (max TM-score to PDB).

As shown in Table~\ref{tab:co-design}, ProteinGenerator and Multiflow consistently show strong performance of structure-sequence compatibility across all sequence lengths, with high scTM scores (up to 0.96±0.06) and relatively low scRMSD values, indicating superior structural quality in generated sequences. ProteinGenerator particularly excels at shorter lengths, showing a balanced performance between quality and diversity metrics. Multiflow maintains high performance even as sequence length increases, demonstrating its robustness with consistently high scTM scores and lower scRMSD values, which indicates its capability to generate high-quality structures. 
ProtPardelle and ESM3, on the other hand, shows degradation in performance with increasing sequence length, as indicated by its low scTM scores and very high scRMSD values, suggesting that it struggles with maintaining structure quality for longer sequences. Overall, these findings suggest that while ProteinGenerator and Multiflow are effective models for generating high-quality protein structures across different lengths, Multiflow is particularly robust across all tested lengths. 
% ProtPardelle and ESM3 have limitations, particularly in handling longer sequences, where their performance is notably weaker.

\subsubsection{Motif-scaffolding}

In this section, we evaluate the performance of various motif-scaffolding methods across different scaffolds used in ~\citet{watson2023rfdiffusion} and \citet{yim2024improved}, focusing on their effectiveness in designing scaffold structures. The primary objective of this evaluation is to compare the efficacy of structure-based and sequence-based approaches in generating designable scaffolds. 
For purely sequence-based methods, e.g., EvoDiff~\citep{alamdari2023protein} and DPLM~\citep{wang2024diffusion}, we use ESMFold to predict the structures of their designed motif-scaffold sequences.

\begin{figure}[h]
    \centering
    \includegraphics[width=\textwidth]{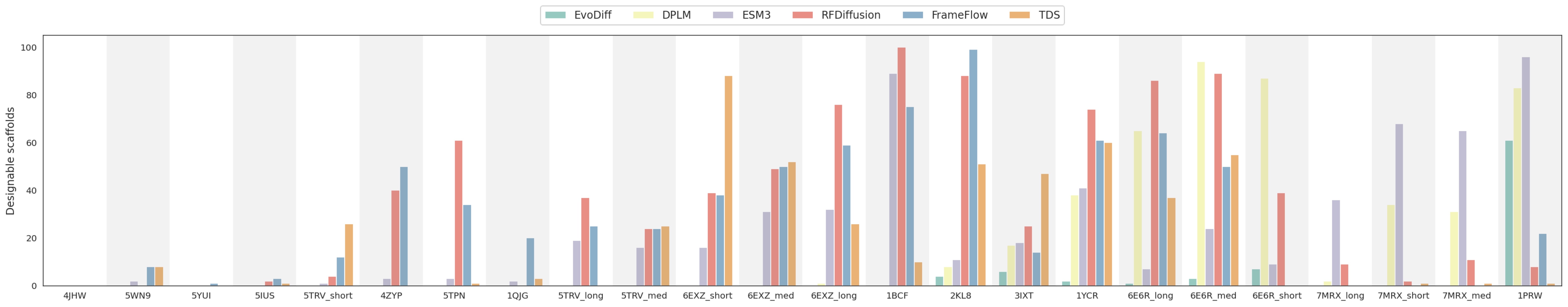}\vspace{-4mm}
    \caption{Performance of motif-scaffolding of structure-based and sequence-based methods on the benchmark used in~\citet{watson2023rfdiffusion}. Results of FrameFlow, RFDiffusion and TDS are quoted from \citet{yim2024improved}\protect\footnotemark.}
    \label{fig:motif-scafold}
\end{figure}
\footnotetext{\url{https://github.com/microsoft/protein-frame-flow/tree/main/motif_scaffolding}} % The text for the footnote

Figure~\ref{fig:motif-scafold} reveals a wide range of performance levels among the tested methods, each exhibiting distinct strengths and weaknesses depending on the specific scaffold context. Notably, structure-based methods such as RFdiffusion~\citep{watson2023rfdiffusion}, TDS~\citep{wu2024practical} and FrameFlow~\citep{yim2024improved} consistently perform well across most scenarios, with RFdiffusion showing particular robustness in generating a high number of designable scaffolds. This suggests that structure-based methods are highly effective at capturing the intricate structural details necessary for successful scaffold design. In contrast, sequence-based methods like EvoDiff and DPLM display variable performance, excelling in certain scaffolds that are primarily governed by evolutionary constraints, but underperforming in others with more complex structural motifs. This variability may reflect their limitations in recognizing and adapting to specific structural features.

Interestingly, ESM3~\citep{hayes2024simulating}, the latest sequence-based method and multimodal language model, capable of perceiving tertiary features through structure tokenization, demonstrates competitive performance in generating designable scaffolds across most cases. Its performance is comparable to that of more advanced structure-based models. This suggests that multimodal language models like ESM3 may effectively integrate structural capabilities within a unified language modeling framework, making them versatile tools for scaffold design. However, ESM3 does not consistently approach structure-based methods across all scenarios, indicating that while multimodal protein language models hold promise, further refinement and optimization are needed to achieve more consistent performance across different structural challenges.

Overall, our findings underscore that no single model currently excels universally across all scaffolds, highlighting the importance of selecting a motif-scaffolding method that aligns with the specific design objectives. Future research should explore the integration of these methods to capitalize on their respective strengths, potentially leading to more robust and versatile scaffold design capabilities

\subsubsection{Antibody design}
\label{sec:antibody_evaluation}

\begin{table}[t]
\footnotesize
\centering
\setlength{\tabcolsep}{2pt}
\renewcommand\arraystretch{1.2}  
\caption{Performance of antibody design methods on 55 antibody-antigen complexes from the RAbD dataset. For methods that can generate multiple antibodies (marked with *), the standard deviations between different antibodies generated against the same antigen are also reported.}
\resizebox{\textwidth}{!}{
\begin{tabular}{llllllll}
%\begin{tabular}{l S[table-format=3.2] S[table-format=3.2] S[table-format=3.2] S[table-format=3.2] S[table-format=3.2] S[table-format=3.2] S[table-format=3.2]}
    \toprule
    & \multicolumn{3}{c}{\textbf{Accuracy}}                                                                                                                      & \multicolumn{1}{c}{\textbf{Functionality}}                         & \multicolumn{3}{c}{\textbf{Specificity}}                                                                                        \\ \cline{2-8} 
    & \AbMethods{AAR} ↑                                & \AbMethods{RMSD} ↓                          & \multicolumn{1}{l|}{\AbMethods{TM-score} ↑}                       & \multicolumn{1}{l|}{\AbMethods{Binding Energy} ↓}                     & \AbMethods{SeqSim-outer} ↓                & \AbMethods{SeqSim-inner} ↑                   & \AbMethods{PHR} ↓                         \\ \cline{2-8} 
    \rowcolor{gray!15} RAbD (natural) & 100.00\%                                         & 0.00                                         & \multicolumn{1}{l|}{1.00}                                         & \multicolumn{1}{l|}{-15.33}                                        & 0.26                                     & N/A                                        & 45.78\%                                 \\
    HERN           & 33.17\%                                       & 9.86                                      & \multicolumn{1}{l|}{0.16}                                      & \multicolumn{1}{l|}{1242.77}                                       & 0.41                            & N/A                                        & \textbf{39.83\%} \\
    MEAN           & 33.47\%                                       & \textbf{1.82}      & \multicolumn{1}{l|}{0.25}                                      & \multicolumn{1}{l|}{263.90}                                         & 0.65                            & N/A                                        & {\ul 40.74\%} \\
    dyMEAN         & \textbf{40.95\%}       & {\ul 2.36*}     & \multicolumn{1}{l|}{{\ul 0.36}}      & \multicolumn{1}{l|}{889.28}                                        & 0.58                                     & N/A                                        & 42.04\%                                 \\
    *dyMEAN-FixFR  & {\ul 40.05\%$\pm$1.06}  & {\color[HTML]{1F2329} 2.37$\pm$0.03}          & \multicolumn{1}{l|}{{\color[HTML]{1F2329} 0.35$\pm$0.01}}          & \multicolumn{1}{l|}{{\color[HTML]{1F2329} 612.75$\pm$56.03}}           & {\color[HTML]{1F2329} 0.60}               & \textbf{0.96}       & {\color[HTML]{1F2329} 43.75\%$\pm$2.24}     \\
    *DiffAb        & {\color[HTML]{1F2329} 35.04\%$\pm$8.36}           & {\color[HTML]{1F2329} 2.53$\pm$0.60}           & \multicolumn{1}{l|}{\textbf{0.37$\pm$0.06}} & \multicolumn{1}{l|}{{\color[HTML]{1F2329} 489.42$\pm$499.76}}          & \textbf{0.37}     & {\color[HTML]{1F2329} 0.45}                & {\color[HTML]{1F2329} 40.68\%$\pm$10.65}    \\
    *AbDPO         & {\color[HTML]{1F2329} 31.29\%$\pm$7.29}           & {\color[HTML]{1F2329} 2.79$\pm$3.01}          & \multicolumn{1}{l|}{{\color[HTML]{1F2329} 0.35$\pm$0.06}}          & \multicolumn{1}{l|}{\textbf{116.06$\pm$186.06}} & {\ul 0.38}    & {\ul 0.60}        & {\color[HTML]{1F2329} 69.69\%$\pm$8.49}     \\
    *AbDPO++       & {\color[HTML]{1F2329} 36.25\%$\pm$7.95}           & {\color[HTML]{1F2329} 2.48$\pm$0.59}          & \multicolumn{1}{l|}{{\color[HTML]{1F2329} 0.35$\pm$0.06}}          & \multicolumn{1}{l|}{{\ul 223.73$\pm$281.7}}  & {\color[HTML]{1F2329} 0.39}              & {\color[HTML]{1F2329} 0.54}                & {\color[HTML]{1F2329} 44.51\%$\pm$9.55}     \\
    \toprule
    & \multicolumn{6}{c}{\textbf{Rationality}}                                                                                                                                                                                                                                                                                   &                                         \\ \cline{2-8}
    & \AbMethods{CN-score} ↑                          & \AbMethods{Clashes-inner} ↓                & \AbMethods{Clashes-outer} ↓                                     & \AbMethods{SeqNat}↑                                                   & \AbMethods{Total Energy} ↓                      & \AbMethods{scRMSD} ↓                        &                                         \\ \cline{2-7}
    \rowcolor{gray!15} RAbD (natural) & 50.19                                        & 0.07                                      & 0.00                                                              & -1.74                                                              & -16.76                                        & 1.77                                      &                                         \\
    HERN          & 0.04                                         & 0.04                                      & 3.25                                                           & \textbf{-1.47}                              & 5408.74                                       & 9.89      &                                         \\
    MEAN           & 1.33                                         & 11.65                                     & 0.29                                                           & -1.83                                                              & 1077.32                                       & 2.77                                      &                                         \\
    dyMEAN         & 1.49                                         & 9.15      & 0.47                                                           & -1.79                                                              & 1642.65                                       & \textbf{2.11}      &                                         \\
    *dyMEAN-FixFR  & {\color[HTML]{1F2329} 1.14$\pm$1.71}             & {\color[HTML]{1F2329} 8.88$\pm$0.55}          & {\color[HTML]{1F2329} 0.48$\pm$0.12}                               & {\color[HTML]{1F2329} -1.82$\pm$0.10}                                   & {\color[HTML]{1F2329} 1239.29$\pm$113.84}         & {\ul 2.48$\pm$0.24} &                                         \\
    *DiffAb        & {\ul 2.02$\pm$2.83}    & {\ul 1.84$\pm$1.35} & {\color[HTML]{1F2329} 0.19$\pm$0.31}                               & {\color[HTML]{1F2329} -1.88$\pm$0.47}                                  & {\color[HTML]{1F2329} 495.69$\pm$350.96}          & {\color[HTML]{1F2329} 2.57$\pm$0.77}          &                                         \\
    *AbDPO         & {\color[HTML]{1F2329} 1.33$\pm$2.31}             & {\color[HTML]{1F2329} 4.14$\pm$1.84}          & {\ul 0.10$\pm$0.24}                       & {\color[HTML]{1F2329} -1.99$\pm$0.34}                                  & \textbf{270.12$\pm$217.45} & {\color[HTML]{1F2329} 2.79$\pm$3.25}          &                                         \\
    *AbDPO++       & \textbf{2.34$\pm$3.20}     & \textbf{1.66$\pm$1.28} & \textbf{0.08$\pm$0.20}                     & {\ul -1.78$\pm$0.43}                         & {\ul 338.14$\pm$266.48} & {\color[HTML]{1F2329} 2.50$\pm$0.75}           &                                        \\
    \bottomrule
\end{tabular}
}%
\label{tab:abtibody_design}
\end{table}
In this section, we selected five antigen-specific antibody design methods (HERN \citep{jin2022antibody}, MEAN \citep{kongconditional}, dyMEAN \citep{kong2023end}, DiffAb \citep{luo2022antigen}, AbDPO \citep{zhou2024antigen}) and two of their variants (dyMEAN-FixFR implemented according to Appendix~\ref{appendix:antibody_design_dyMEANFixFR} and AbDPO++), making a total of seven methods, to evaluate their performance in CDR-H3 generation towards the given antigens. All methods were trained on the same dataset with parameters reported in the corresponding papers and tested on a common set of 55 test cases from the RAbD dataset \citep{adolf2018rosettaantibodydesign}, details refer to Appendix~\ref{appendix:antibody_design_dataset}. Notably, dyMEAN-FixFR is not an official variant of dyMEAN; we modified dyMEAN to align its task setting with the other methods and allow it to generate different antibodies for the same antigen. The final evaluation results are shown in Table~\ref{tab:abtibody_design}. For each evaluation metric, we highlighted the \textbf{best} performance in bold and the \underline{second-best} with the underline, the detailed implementation of each metric could be seen at Appendix~\ref{appendix:antibody_design_metrics}.

In the \textbf{Accuracy} evaluation, dyMEAN and MEAN achieved the best performance in terms of sequence and structure (highest \AbMethods{AAR} and lowest \AbMethods{RMSD}), while DiffAb performed best in \AbMethods{TM-score}. However, considering multiple evaluation metrics, these methods did not perform as well overall. Additionally, apart from HERN, there were no significant performance differences among the other methods.

In the evaluation of \textbf{Functionality}, all methods produced antibodies with binding energies to the given antigens significantly higher than those of natural antibodies. AbDPO and AbDPO++ achieved the best performance among all methods by aligning on binding energy.

In the \textbf{Specificity} evaluation of antibodies, we mainly observed the sequence similarity between antibodies against different antigens (\AbMethods{SeqSim-outer}) and the proportion of hydrophobic residues in the generated antibodies (\AbMethods{PHR}). The former metric indicates whether the method can design antibodies specific to a given antigen, while the latter reflects the potential non-specific binding due to high hydrophobicity.

% \begin{xQuote}
\begin{itemize}[leftmargin=*]
  \item In \AbMethods{SeqSim-outer}, we noted that MEAN and dyMEAN generated highly similar sequences for different antigens (the maximum \AbMethods{SeqSim-outer} in our test set was 0.79, indicating that all antibody differences came only from length variations). This suggests that their excellent AAR might stem from learning high-frequency patterns in antibody sequences, generating antibodies according to these patterns for different antigens. In contrast, DiffAb and AbDPO performed the best. 
  
  \item For methods that can generate different antibodies for the same antigen, we also measured the sequence similarity among different antibodies generated for the same antigen (\AbMethods{SeqSim-inner}). We expect antibodies generated for the same antigen to be more similar. In this aspect, dyMEAN-FixFR and AbDPO performed the best. However, the 0.96 \AbMethods{SeqSim-inner} of dyMEAN-FixFR indicates that despite introducing randomness during model initialization, the final sequence generation showed almost no differences. Additionally, DiffAb, which performed best in \AbMethods{SeqSim-outer}, generated less similar antibodies for the same antigen, suggesting possible underfitting in sequence generation. Considering both types of \AbMethods{SeqSim}, AbDPO achieved the best performance. 
  
  \item In \AbMethods{PHR}, HERN and dyMEAN performed the best, but overall, almost all methods performed better than natural antibodies. Only AbDPO generated an excessive number of hydrophobic residues, reducing specificity. However, its variant, AbDPO++, controlled \AbMethods{PHR} well, closely matching natural antibodies among all methods.

\end{itemize}
% \end{xQuote}

The \textbf{Rationality} evaluation includes three aspects: structural rationality, sequence rationality, and joint structural and sequence rationality. 

% \begin{xQuote}
\begin{itemize}[leftmargin=*]
  \item In structural rationality, we focused on the score for peptide bond lengths conforming to the natural peptide bonds length distribution (\AbMethods{CN-score}), the number of potential internal clashes in the generated structure (\AbMethods{Clashes-inner}), and the clashes between the generated structure and other parts (\AbMethods{Clashes-outer}). It was evident that irrational structures were prevalent in generated antibodies, but overall, diffusion-based methods performed better. AbDPO++ and DiffAb achieved the best performance among all methods. HERN and MEAN/dyMEAN exhibited different tendencies in \AbMethods{Clashes-inner/outer}, corresponding to our observations of the generated samples. HERN tends to generate large CDR-H3 structures, leading to fewer internal clashes but more clashes with the antigen, whereas MEAN/dyMEAN tends to generate smaller CDR-H3 structures. 
  
  \item In sequence rationality, we used the inverse perplexity of AntiBERTy \citep{ruffolo2021deciphering} to represent sequence naturalness, \AbMethods{SeqNat}, showing that HERN performed the best, possibly due to HERN being the only auto-regressive model. AbDPO++ achieved the second-best performance and was closest to natural antibodies. 
  
  \item In the joint evaluation of structure and sequence, we mainly focused on the consistency between the generated structure and sequence from two perspectives: physical energy and structure prediction. In terms of physical energy, we calculated the total energy of the generated CDR-H3s (\AbMethods{Total Energy}), which would be severely affected by the clashes caused by sidechains and thus reflect the irrationality between the generated structure and sequence. In this energy-related metric, AbDPO and AbDPO++ performed best among all methods. From the perspective of structure prediction, we used IgFold \citep{ruffolo2023fast} to predict the structure of the generated sequence, performed a post-optimization with the antigen as the condition, and calculated the CA-RMSD between the predicted structure and the generated structure (\AbMethods{scRMSD}). dyMEAN and dyMEAN-FixFR performed best in \AbMethods{scRMSD}. Although these two metrics both reflect the consistency between sequence and structure, they focus on different aspects. Moreover, both energy calculations and structure predictions have inherent errors, so the performance of different methods may not be consistent across these two metrics.
\end{itemize}
% \end{xQuote}

Overall, evaluating antibody design methods encompasses various aspects, and using only a few metrics will seriously mislead researchers' understanding of model performance. Moreover, we must recognize that no single method outperformed all others across the board, and all methods showed substantial gaps compared to natural antibodies. 
The discrepancies may come from the severe lack of structured data, causing models to focus on certain sequence patterns or structures. Additionally, most models do not perform atomic-level modeling of antibodies and antigens, preventing accurate interaction modeling. New task paradigms must be developed to overcome current challenges in antibody design.
%A key factor may be the scarcity of available structural data and new task paradigms should be developed. 
Nonetheless, AbDPO++, by utilizing synthetic data and aligning with various properties, achieved one of the best performances in almost all aspects among all methods, without exhibiting obvious weaknesses.

\subsection{Protein Conformation Prediction}
% In the second part of ProteinBench, we focus on conformation prediction, another class of cross-modality tasks aimed at predicting protein structures (conformations) from their sequences. While the current models are based on a body of work distinct from the design tasks, the ability to predict protein conformations provides insight into a model's understanding of the physics and dynamics of protein structures. This capability is essential for future protein foundation models to fully understand, predict, and design proteins that embody the key sequence-structure-function relationships

Protein conformation prediction infers the 3D \textit{conformations} of proteins from their sequences, evaluating models based on their understanding of structure, dynamics, and ultimately functions. 
We begin by benchmarking common folding models, as they play a crucial role in the development of conformation prediction models. We then compare five recent studies that explore difference strategies to extend folding to conformation prediction, focusing on \textit{multiple-state} and \textit{distribution prediction} tasks. Key differences between these approaches are outlined in Appendix \ref{apdx:model_compare}.

% To date, a comprehensive comparison of these methods has not been conducted, making this the first benchmark study to evaluate and compare these strategies.

% [Give some examples on the connection of conformation/dynamics and design. ]
% Predicting protein conformation has been challenging due to limited data. Recent works have explored several strategies: (1) perturbing the sequence input of folding models \citep{delalamoSamplingAlternativeConformational2022,wayment2024predictingMSA}; (2) perturbing protein structures \citep{luStr2StrScorebasedFramework2024}; (3) training generative models on large-scale structural data from experiments or simulations \citep{jingEigenFoldGenerativeProtein2023,jingAlphaFoldMeetsFlow2023,wang2024proteinconfdiff,zheng2024predictingDiG}; (4) improving the conformational sampling using physical models \citep{zheng2024predictingDiG,wang2024proteinconfdiff}. 

\subsubsection{Protein Folding: single-state prediction}

As shown in Table \ref{tab:folding_cameo2022}, Multiple Sequence Alignment (MSA) based folding models (AlphaFold2, OpenFold, RoseTTAFold2) outperforms protein language model-based approaches (ESMFold, EigenFold). While the predicted structure quality is comparable among the all-atom resolution models, AlphaFold2 and its reproduction, OpenFold, achieve the best performance across all accuracy metrics, offering a strong foundation for conformation prediction.
\begin{table}[t]\footnotesize
% \vspace{-3mm}
  \centering\setlength{\tabcolsep}{2pt}
  \caption{Performance of protein folding on CAMEO2022. Results are reported as mean/median over 183 proteins. The \textbf{best} performance is highlighted in bold, and the \uline{second-best} is underlined. EigenFold only predicts CA coordinates and PepBond break \% is not available (shown in ``N/A'').}
\resizebox{0.8\textwidth}{!}{%

% Top 1 only
\begin{tabular}{lcccccc}
\toprule
& \multicolumn{4}{c}{\textbf{Accuracy}} & \multicolumn{2}{c}{\textbf{Quality}} \\
 \cmidrule(lr){2-5} \cmidrule(lr){6-7}
& \textbf{TM-score} ↑ & \textbf{RMSD} ↓ & \textbf{GDT-TS} ↑ & \textbf{lDDT} ↑ & \textbf{CA clash} (\%) ↓ & \textbf{PepBond break} (\%) ↓ \\
\midrule   
AlphaFold2 & \textbf{0.871}/\textbf{0.952} & \textbf{3.21}/\uline{1.64} & \textbf{0.860}/\textbf{0.921} & \textbf{0.904}/\textbf{0.933} & \textbf{0.3}/\textbf{0.0} & 4.8/4.1 \\
OpenFold & \uline{0.870}/\uline{0.947} & \textbf{3.21}/\textbf{1.59} & \uline{0.856}/\uline{0.913} & \uline{0.899}/\textbf{0.933} & 0.4/\textbf{0.0} & \textbf{2.0}/\textbf{1.7} \\
RoseTTAFold2 & 0.859/0.941 & 3.52/1.75 & 0.845/0.903 & 0.892/0.926 & \textbf{0.3}/\textbf{0.0} & 5.5/4.0 \\
ESMFold & 0.847/0.929 & 3.98/2.10 & 0.826/0.881 & 0.870/0.907 & \textbf{0.3}/\textbf{0.0} & \uline{4.7}/\uline{3.4} \\
EigenFold & 0.743/0.823 & 7.65/3.73 & 0.703/0.781 & 0.737/0.810 & 8.0/4.6 & N/A \\
\bottomrule
\end{tabular}

}%
\label{tab:folding_cameo2022}%
\end{table}%

\subsubsection{Multiple-state prediction}

We first investigate the results on \textbf{BPTI} with the best model of each study highlighted in Table \ref{tab:conf_bpti_simple} and the complete evaluation in Table \ref{tab:conf_bpti}. 
% ConfDiff model with force-guidance showed the best overall accuracy (RMSDens) on predicting the five known metastable states, while 
% Detailed comparisons revealed additional insights. 
The classifier-free guidance in ConfDiff~\citep{wang2024proteinconfdiff} achieved performance comparable to fine-tuning on MD conformation data and, when combined with force guidance, delivered the best overall accuracy (RMSDens). This suggests that incorporating structural exploration and physical constraints can enhance the sampling of high-accuracy conformations. However, structural exploration alone may be error-prone as Str2Str~\citep{luStr2StrScorebasedFramework2024}, the structure-only models, showed low accuracy even with small perturbations. Other strategies to enhance conformation sampling also showed improved performance: MSA subsampling~\citep{delalamoSamplingAlternativeConformational2022} with reduced MSA depth excelled at sampling Cluster 3, the most difficult state to capture, and ESMFlow~\citep{jingAlphaFoldMeetsFlow2023} fine-tuned on the MD dataset showed improved diversity and accuracy compared to the PDB-trained base model. However, these approaches also experienced a decline in quality, with increased CA clashing or peptide bond breaking. Lastly, for most methods, increasing the sample depth ($N$) significantly improved expected accuracy, suggesting that a sufficient sample size is essential for thorough evaluation.

% EigenFold also shows lower diversity, which may limit its effectiveness in sampling diverse conformations. 
% \input{tables/conf_bpti_orig}
\begin{table}[t]\footnotesize
  \centering\setlength{\tabcolsep}{2pt}
  \caption{Performance of the multiple-state prediction on BPTI. \textit{Accuracy} metrics (RMSDens, RMSD Cluster 3) are reported as the mean and standard deviations from 20 bootstrap samples, at different sample sizes. \textit{Diversity} and \textit{quality} are evaluated based on 1,000 conformations.}

\resizebox{\textwidth}{!}{%

\begin{tabular}{lcccccccccc}
\toprule
 & \multicolumn{3}{c}{\textbf{RMSDens (Å)} ↓} & \multicolumn{3}{c}{\textbf{RMSD (Å) Cluster 3} ↓} & \textbf{Diversity} & \multicolumn{2}{c}{\textbf{Quality}} \\
 \cmidrule(lr){2-4}  \cmidrule(lr){5-7} \cmidrule(lr){8-8}  \cmidrule(lr){9-10}
 & \textbf{N=10} & \textbf{N=100} & \textbf{N=1000} & \textbf{N=10} & \textbf{N=100} & \textbf{N=1000} & \textbf{\specialcell{Pairwise\\RMSD}} & \textbf{CA} \textbf{clash\%} ↓ & \textbf{PepBond} \textbf{break\%}↓ \\
\midrule

EigenFold & 1.56$\pm$0.02 & 1.50$\pm$0.01 & 1.46$\pm$0.00 & 2.54$\pm$0.03 & 2.48$\pm$0.01 & 2.46$\pm$0.01 & 0.85 & 1.4 & N/A \\
MSA-depth32 & \uline{1.66$\pm$0.03} & 1.54$\pm$0.04 & 1.41$\pm$0.02 & \textbf{2.43$\pm$0.06} & \textbf{2.19$\pm$0.16} & \textbf{1.85$\pm$0.05} & 2.14 & 0.6 & \uline{10.6} \\
Str2Str-ODE ($T_\text{max}=0.15$) & 2.40$\pm$0.12 & 2.20$\pm$0.05 & 2.09$\pm$0.01 & 3.00$\pm$0.20 & 2.73$\pm$0.12 & 2.58$\pm$0.05 & 1.86 & \textbf{0.0} & 13.9 \\
ESMFlow-MD & 1.68$\pm$0.06 & \uline{1.47$\pm$0.04} & \uline{1.39$\pm$0.03} & \uline{2.44$\pm$0.11} & \uline{2.27$\pm$0.10} & \uline{2.18$\pm$0.02} & 1.17 & \textbf{0.0} & 14.3 \\
ConfDiff-ESM-Force & \textbf{1.58$\pm$0.04} & \textbf{1.43$\pm$0.03} & \textbf{1.36$\pm$0.01} & \uline{2.44$\pm$0.06} & 2.35$\pm$0.05 & 2.24$\pm$0.06 & 1.76 & 0.1 & \textbf{8.9} \\
\bottomrule
\end{tabular}

}%

\label{tab:conf_bpti_simple}%
\end{table}%

\begin{table}[t]\footnotesize
% \vspace{-1mm}
  \centering\setlength{\tabcolsep}{2pt}
  \caption{Performance of multiple-state prediction on \textit{apo-holo}. \textit{apo}/\textit{holo}-TM represents the maximum TM-score of the samples relative to the reference \textit{apo}/\textit{holo} structure. 20 conformations were sampled for each protein, and the results are reported as mean across 91 proteins.}
\resizebox{0.8\textwidth}{!}{%

% No TMmin
\begin{tabular}{lcccccc}
\toprule
 & \multicolumn{3}{c}{\textbf{Accuracy}} & \textbf{Diversity} & \multicolumn{2}{c}{\textbf{Quality}} \\
  \cmidrule(lr){2-4}  \cmidrule(lr){5-5} \cmidrule(lr){6-7}
 & \textbf{\textit{apo}-TM} ↑ & \textbf{\textit{holo}-TM} ↑  & \textbf{TMens} ↑ & \textbf{Pairwise TM} & \textbf{CA clash \%} ↓ & \textbf{\specialcell{PepBond break \%}} ↓ \\
 
\midrule
\rowcolor{gray!15}
\textit{apo} model & 1.000 & 0.790 & 0.895 & N/A & N/A & N/A \\
EigenFold & 0.831 & 0.864 & 0.847 & 0.907 & 3.6 & N/A \\
MSA-depth256 & 0.845 & \uline{0.889} & \uline{0.867} & 0.978 & \textbf{0.2} & \textbf{4.6} \\
Str2Str-ODE ($T_\text{max}=0.3$) & 0.766 & 0.781 & 0.774 & 0.872 & \textbf{0.2} & 14.7 \\
AlphaFlow-PDB & \textbf{0.855} & \textbf{0.891} & \textbf{0.873} & 0.924 & 0.3 & 6.6 \\
ConfDiff-Open-PDB & \uline{0.847} & 0.886 & \uline{0.867} & 0.909 & 0.5 & \uline{5.5} \\

\bottomrule
\end{tabular}

}%

\label{tab:conf_apo_simple}%
\end{table}%

\textbf{\textit{apo-holo}} is a more challenging dataset where models are required to predict both the unbound (\textit{apo}) and bound (\textit{holo}) conformations induced by ligand binding. As shown in Table~\ref{tab:conf_apo_simple} and Table~\ref{tab:conf_apo}, strategies to enhance conformation diversity did not improve the TMens score, and the best-performing models closely resemble folding models. In comparison, a baseline model that consistently predicts the perfect \textit{apo} structure achieved a higher TMens score. These findings suggest that these models struggle to accurately sample \textit{apo-holo} conformation changes, and higher accuracy may stem from using a stronger folding model.

In summary, strategies like MSA subsampling, guidance during diffusion, or training on MD conformation data can improve sample diversity and enhance ensemble accuracy for certain local dynamics (as in BPTI). However, they may not be sufficient to capture the complex dynamics involved in processes like \textit{apo}-\textit{holo} conformational changes.

\subsubsection{Distribution prediction}

% In this final task, we benchmark models on the ATLAS test set, which includes 82 proteins, and focus on each model's ability to recover the conformational distribution observed in classic protein molecular dynamics simulations. The results are summarized in Table \ref{tab:conf_atlas}. For comparison, we include reference performances of (1) i.i.d. samples (MD iid) from reference MD trajectories and (2) 250 consecutive samples, corresponding to 2.5 ns of simulation time (MD 2.5 ns).

In this section, we focus on predicting the conformational distributions observed in classic molecular dynamics simulations. The results are summarized in Table \ref{tab:conf_atlas_simple} and Table \ref{tab:conf_atlas}. We include reference performances of (1) MD iid: i.i.d. samples from reference MD trajectories and (2) MD 2.5 ns: consecutive samples from the trajectories corresponding to 2.5 ns of simulation.

Overall, generative models trained to sample protein conformations (AlphaFlow, ConfDiff) significantly outperform perturbation-based methods (MSA subsampling and Str2Str), regardless of perturbation levels. Using a strong folding model like AlphaFold2 generally results in higher accuracy. Classifier-free guidance in ConfDiff improved distribution sampling but were less effective than direct fine-tuing on MD data, highlighting the importance of large-scale conformational data for future models. Additionally, the trade-offs between diversity, prediction performance, and sample quality persist: fine-tuning on MD data improves sample diversity and performance for AlphaFlow but decreases sample quality. 

AlphaFlow and ConfDiff models fine-tuned on MD data have shown promise in capturing the conformational distributions, achieving performance comparable to that of 2.5 ns MD simulations on some metrics. However, a key goal for these models is to achieve i.i.d. sampling equivalent to long-term MD simulations, and benchmark results reveal a remaining gap in reaching this objective.

\begin{table}[t]\footnotesize
% \vspace{-3mm}
  \centering\setlength{\tabcolsep}{2pt}
  \caption{Performance on distribution prediction for ATLAS. 250 conformations were sampled for each protein and the median values across 82 proteins are reported. \textit{*These metrics are not available for models that lack side-chain or full backbone information}.}

\resizebox{\textwidth}{!}{%
\begin{tabular}{lccccccccc}
\toprule
 & \multicolumn{2}{c}{\textbf{Diversity}} & \multicolumn{3}{c}{\textbf{Flexibility: \textit{Pearson} $r$ on}} & \multicolumn{4}{c}{\textbf{Distributional accuracy}} \\
 \cmidrule(lr){2-3}  \cmidrule(lr){4-6} \cmidrule(lr){7-10}
 & \textbf{\specialcell{Pairwise\\RMSD}} & \textbf{*RMSF} & \specialcell{\textbf{Pairwise}\\\textbf{RMSD} ↑ } & \specialcell{*\textbf{Global}\\\textbf{RMSF} ↑ }  & \specialcell{*\textbf{Per target}\\\textbf{RMSF} ↑ } & \specialcell{*\textbf{RMWD} ↓} & \specialcell{\textbf{MD PCA}\\\textbf{W2} ↓} & \specialcell{\textbf{Joint}\\\textbf{PCA W2} ↓} & \specialcell{\textbf{PC sim}\\\textbf{$>$ 0.5 \%}↑ } \\
\midrule
\rowcolor{gray!15}
MD iid & 2.76 & 1.63 & 0.96 & 0.97 & 0.99 & 0.67 & 0.73 & 0.71 & 93.9 \\
\rowcolor{gray!15}
MD 2.5ns & 1.54 & 0.98 & 0.89 & 0.85 & 0.85 & 2.22 & 1.55 & 1.89 & 36.6 \\
EigenFold & 5.96 & N/A & -0.03 & N/A & N/A & N/A & 2.31 & 7.96 & 12.2 \\
MSA-depth256 & 0.83 & 0.53 & 0.25 & 0.34 & 0.59 & 3.60 & 1.79 & 2.91 & 29.3 \\
Str2Str-ODE ($T_\text{max}=0.1$) & 1.66 & N/A & 0.13 & N/A & N/A & N/A & 2.14 & 4.39 & 6.1 \\
AlphaFlow-MD & 2.87 & 1.63 & \uline{0.53} & \uline{0.66} & \textbf{0.85} & \textbf{2.64} & \uline{1.55} & \uline{2.29} & \textbf{39.0} \\
ConfDiff-Open-MD & 3.43 & 2.21 & \textbf{0.59} & \textbf{0.67} & \textbf{0.85} & \uline{2.75} & \textbf{1.41} & \textbf{2.27} & \uline{35.4} \\
\bottomrule
 & \multicolumn{4}{c}{\textbf{Ensemble observables}} & \multicolumn{2}{c}{\textbf{Quality}}  \\
  \cmidrule(lr){2-5}  \cmidrule(lr){6-7}
 & \specialcell{\textbf{Weak}\\\textbf{contacts $J$} ↑} & \specialcell{\textbf{Transient}\\\textbf{contacts $J$}↑} & \specialcell{*\textbf{Exposed} \\\textbf{residue $J$} ↑} & \specialcell{*\textbf{Exposed MI} \\\textbf{matrix $\rho$} ↑} & \specialcell{\textbf{CA clash}\\\textbf{\%} ↓} & \specialcell{*\textbf{PepBond}\\\textbf{break \%} ↓} &  & &  \\
\midrule
\rowcolor{gray!15} 
MD iid & 0.90 & 0.80 & 0.93 & 0.56 & 0.0 & 3.4 \\
\rowcolor{gray!15} 
MD 2.5ns & 0.62 & 0.45 & 0.64 & 0.25 & 0.0 & 3.4 \\
EigenFold & 0.36 & 0.19 & N/A & N/A & 5.6 & N/A \\
MSA-depth256 & 0.30 & 0.29 & 0.36 & 0.06 & \textbf{0.0} & \textbf{5.5} \\
Str2Str-ODE ($T_\text{max}=0.1$) & 0.42 & 0.18 & N/A & N/A & \textbf{0.0} & 12.1 \\
AlphaFlow-MD & \uline{0.62} & \textbf{0.41} & \textbf{0.69} & \textbf{0.35} & \textbf{0.0} & 22.2 \\
ConfDiff-Open-MD & \textbf{0.63} & \uline{0.39} & \uline{0.65} & \uline{0.33} & 0.5 & \uline{6.5} \\
\bottomrule
\end{tabular}
}%
\label{tab:conf_atlas_simple}%
\end{table}%j

\section{Conclusions and Future Work}

In summary, we present the first comprehensive study evaluating the capabilities of various protein foundation models across eight distinct tasks, with a particular focus on protein design and conformation dynamics. We have developed a unified, multi-metric evaluation framework, which is essential for unbiased assessment of protein foundation models from multiple facets. Based on the performance results, we provide insights and considerations for the development and effective use of protein foundation models, offering guidance for future research. We highlight the key observations from our holistic evaluation as follows.

\subsection{Key observations}
\textbf{Valid evaluation of protein foundation models necessitates the use of correct and comprehensive evaluation metrics.} The emergence of advanced folding models, exemplified by AlphaFold2 and ESMFold, has opened up valuable opportunities for accurately assessing the quality, stability, and precision in protein generative tasks. However, it is crucial to acknowledge that, due to their current limitations in complex structure prediction capabilities, \textbf{certain tasks may still lack sufficiently accurate evaluation methods}. For example, within the realm of antibody design, researchers have at times been misled by reconstruction metrics like Amino Acid Recovery (AAR) and Root Mean Square Deviation (RMSD) related to accuracy, resulting in overly optimistic conclusions. Here, we intend to tackle this challenge by introducing a combined evaluation approach, integrating reconstitution and physical rationality metrics. Also, considering the inherent complexity of protein scientific problems, it becomes imperative to adopt a \textbf{multifaceted evaluation strategy} to capture various facets of protein structure and function. Here, in ProteinBench, we aims to capture various facets of protein structure and function, fostering a more holistic understanding of the performance of foundation models in protein-related tasks. Furthermore, \textbf{metrics alone are insufficient}. In the development of generative models for protein, the primary objective is to accurately fit the distribution of the training data. Our evaluation approach extends beyond simple comparisons of metric values. We have implemented a more comprehensive assessment strategy that includes measuring the same metrics for the training data (which encompasses native proteins, antibodies, and molecular dynamics conformations in various lengths). This provides a high-resolution gold reference for protein generative targets, allowing for a more contextually rich evaluation framework. 
% \dongyu{In antibody design task, researchers have been misled by \AbMethods{Accuracy}-related metrics (like AAR and RMSD), leading to an overly optimistic view of the results. All methods exhibit significant discrepancies when evaluated at a fine atomic level or energy level compared to natural antibodies. The underlying factors contributing to these issues include the extreme scarcity of data and the insufficiencies in modeling protein interactions at the atomic level.}

\textbf{No single model currently excels across all protein design objectives. The choice of model should be carefully aligned with the intended applications.} In the field of protein foundation models, two primary approaches have emerged: language models and geometric models. Each approach has its strengths and limitations, which are reflected in the performance of ProteinBench. We found language models show good performance in capturing nature evolution distributions. This is evidenced by their high accuracy in native sequence recovery (inverse-folding) and high quality in scaffolding evolution-conserved motifs. However, language models show limitations in robustness when designing sequence for de novo backbones, and in generating novel sequences for sequence-based protein design. In contrast, structure-based models exhibit greater robustness and tolerance for structural noises in de novo design task, and show greater potential for creating proteins with new folds or functions. These findings underscore the importance of carefully considering specific design objectives when researchers are selecting a model to use.

\textbf{While generative models extended from classic folding models have shown ability to sample protein conformations, challenges remain in both multiple-state prediction and distribution prediction.} 
Protein conformation prediction is a new but crucial assessment of the multi-modal capabilities and physical understanding of protein foundation models. While strategies proposed in current models may benefit certain tasks, they often provide limited improvement in others. For example, although fine-tuning models using the MD conformation dataset showed promising results on the ATLAS benchmark, little to no improvement was observed in the multi-state prediction of \textit{apo}-\textit{holo} conformations. Additionally, the common trade-off between diversity and quality in current models underscores the importance of consistent evaluation across the dimensions of accuracy, diversity, and quality in protein conformation prediction tasks.

\subsection{Limitations and Future Work}
 We acknowledge several limitations and opportunities for enhancement in our current benchmark: (1) The selection of foundation models may not be exhaustive. Future iterations should incorporate additional foundation models to provide a more comprehensive comparison. (2) Inconsistencies in training data across models currently hinder direct comparisons of different model architectures. Future work could address this by standardizing datasets, allowing for more accurate comparisons of architectural performance. (3) The benchmark could be expanded to include a wider range of tasks, further broadening its scope and utility. We are committed to continually refining and expanding ProteinBench. Our vision is for it to evolve into a dynamic, growing benchmark that accelerates progress in the field of protein modeling and design.

\section*{Acknowledgements}
This benchmark represents a collaborative effort from our research group, with each member contributing significantly from their respective areas of expertise. The diverse insights and analyses provided by each contributor have been instrumental in shaping this comprehensive work. Q. Gu conceived and oversees the project. F. Ye coordinated the experiments and analysis, while also conducting model evaluations for inverse folding, backbone design, and a portion of the sequence design tasks. Z. Zheng was responsible for model evaluations and results analysis in sequence design, co-design, and motif scaffolding. D. Xue carried out model evaluation and metrics analysis for antibody design. Y. Shen and L. Wang conducted model evaluation and analysis for single and multiple-state prediction as well as conformation distribution prediction. 
F. Ye initiated and drove the writing of the paper with contributions from all other authors. 
We are grateful for the dedication and expertise demonstrated by each team member. Their collective efforts have been crucial in developing this multifaceted benchmark.

\appendix
\section{Overview of protein foundation model benchmarks}\label{appendix:sum_benchmarks}

In this section, we provide a comprehensive overview of existing benchmarks for protein foundation models. Table \ref{tab:benchmarks} illustrates the current landscape of these benchmarks, revealing significant limitations in the scope and applicability. The majority of existing benchmarks are narrowly focused, primarily addressing task-specific evaluations rather than offering a holistic assessment of protein foundation models.

The benchmarks we examined can be broadly categorized into two main groups: those focused on protein design tasks and those evaluating protein conformational dynamics. Within the protein design category, we observe that while inverse folding is well-represented across multiple benchmarks, other crucial aspects such as backbone design, sequence design, and structure-sequence co-design are often overlooked. Similarly, in the realm of protein conformational dynamics, only a few benchmarks address critical tasks like single-state and multiple-state prediction.

Notably, our proposed ProteinBench stands out by offering the most comprehensive coverage across various tasks. It encompasses a wide range of evaluations, including inverse folding, backbone design, sequence design, structure-sequence co-design, and antibody design in the protein design domain, as well as single-state folding, and multiple-state prediction in the conformational dynamics domain.
% Table generated by Excel2LaTeX from sheet 'New binder'

\begin{table}[htbp]\footnotesize
  \setlength{\tabcolsep}{2pt}
  \centering
  \caption{A comparison of benchmarks for protein fundamental tasks.}
  \resizebox{\textwidth}{!}{%
    \begin{tabular}{lccccccccc}
    \toprule
    \textbf{Benchmark}  & \multicolumn{6}{c}{\textbf{Protein Design}}  & \multicolumn{3}{c}{\textbf{Protein Conformation Prediction}}\\
    \midrule
    & \begin{tabular}[c]{@{}c@{}}Inverse\\ Folding\end{tabular} & 
      \begin{tabular}[c]{@{}c@{}}Backbone\\ Design\end{tabular} & 
      \begin{tabular}[c]{@{}c@{}}Sequence\\ Design\end{tabular} & 
      \begin{tabular}[c]{@{}c@{}}Struc-Seq\\ Codesign\end{tabular} & 
      \begin{tabular}[c]{@{}c@{}}Motif\\ scaffolding\end{tabular} & 
      \begin{tabular}[c]{@{}c@{}}Antibody\\ Design\end{tabular} & 
      \begin{tabular}[c]{@{}c@{}}Folding\\ (single-state)\end{tabular} & 
      \begin{tabular}[c]{@{}c@{}}Multiple State\\ Prediction\end{tabular} & 
      \begin{tabular}[c]{@{}c@{}}Distribution\\ Prediction\end{tabular} \\
    \midrule

    PDB-Struct~\citep{wang2023pdb} & \color{green}{\cmark} & \color{red}{\xmark} & \color{red}{\xmark} & \color{red}{\xmark} & \color{red}{\xmark} &  \color{red}{\xmark} & \color{red}{\xmark} & \color{red}{\xmark} & \color{red}{\xmark}\\
    Proteininvbench~\citep{gao2024proteininvbench}  & \color{green}{\cmark} & \color{red}{\xmark} & \color{red}{\xmark} & \color{red}{\xmark} & \color{red}{\xmark} &  \color{red}{\xmark} & \color{red}{\xmark} & \color{red}{\xmark} & \color{red}{\xmark}\\
    RFDiffusion~\citep{watson2023rfdiffusion}  & \color{red}{\xmark} & \color{red}{\xmark} & \color{red}{\xmark} & \color{red}{\xmark} & \color{green}{\cmark}&  \color{red}{\xmark} & \color{red}{\xmark} & \color{red}{\xmark} & \color{red}{\xmark}\\
    CASP~\citep{casp15}  & \color{red}{\xmark} & \color{red}{\xmark} & \color{red}{\xmark} & \color{red}{\xmark} &  \color{red}{\xmark} & \color{red}{\xmark} & \color{green}{\cmark} & \color{red}{\xmark} & \color{red}{\xmark}\\
    CAMEO~\citep{Robin2021} & \color{red}{\xmark} & \color{red}{\xmark} & \color{red}{\xmark} & \color{red}{\xmark} &  \color{red}{\xmark} & \color{red}{\xmark} & \color{green}{\cmark} & \color{red}{\xmark} & \color{red}{\xmark}\\
    ProteinBench & \color{green}{\cmark} & \color{green}{\cmark}  & \color{green}{\cmark}  & \color{green}{\cmark}  & \color{green}{\cmark}  &  \color{green}{\cmark}  & \color{green}{\cmark}  & \color{green}{\cmark} & \color{green}{\cmark}\\
    \bottomrule
    \end{tabular}%
    }%
  \label{tab:benchmarks}%
\end{table}%

\section{Additional Details on Benchmarking Evaluations}
\subsection{Additional results for protein design}
% Table generated by Excel2LaTeX 
\begin{table}[t]\footnotesize
  \centering\setlength{\tabcolsep}{2pt}
  \caption{Performance of backbone design models evaluated using 200, and 400 amino acids in lengths. The reported results are the medium of repetitive experiments. We use bold text to highlight the best and suboptimal results for each metric. For the novelty and diversity metrics, we only highlight results with the corresponding scTM score higher than 0.5. 'N/A' stands for not applicable.}
    \resizebox{\textwidth}{!}{%
    \begin{tabular}{lccccc|ccccc|}
    \toprule
      & \multicolumn{5}{c}{\textbf{length 200}} & \multicolumn{5}{c}{\textbf{length 400}}  \\

    \cmidrule{2-11}  & \multicolumn{2}{c}{\textbf{Quality}} & \multicolumn{1}{c}{\textbf{Novelty}}  & \multicolumn{2}{c}{\textbf{Diversity}}  & \multicolumn{2}{c}{\textbf{Quality}} & \multicolumn{1}{c}{\textbf{Novelty}}  & \multicolumn{2}{c}{\textbf{Diversity}} \\

    \cmidrule{2-11}  & \multicolumn{1}{c}{\textbf{scTM ↑}} & \multicolumn{1}{c}{\textbf{scRMSD ↓}}  & \multicolumn{1}{c}{\textbf{Max TM ↓}} & \multicolumn{1}{c}{\textbf{pairwise TM ↓}} & \multicolumn{1}{c}{\textbf{Max Clust. ↑}}  & \multicolumn{1}{c}{\textbf{scTM ↑}} & \multicolumn{1}{c}{\textbf{scRMSD ↓}}  & \multicolumn{1}{c}{\textbf{Max TM ↓}} & \multicolumn{1}{c}{\textbf{pairwise TM ↓}} & \multicolumn{1}{c}{\textbf{Max Clust.↑}} \\
    \midrule

    Native PDBs  & 0.974 &	0.674 &	N/A &	0.278 & 0.790 & 0.970 &	1.085 &  N/A &	0.261 &	0.840 \\
    \midrule
    RFdiffusion  & 0.982 & 0.617 & 0.638 & 0.363 & 0.64 & 0.927 &	2.12 & 0.634 & 0.356 & 0.720 \\
    FrameFlow    & 0.982 &	0.617 & 0.638 & 0.363 & 0.640 & 0.927 &	2.12 &  0.634 & 0.356 & 0.720 \\
    Chroma       & 0.892 & 1.776 & 0.674 & 0.346 & 0.620 & 0.761 & 4.891 & 0.626 & 0.304 & 0.95 \\
    FrameDiff (latest) & 0.893 & 1.789 & 0.689 & 0.464 & 0.260 & 0.800 & 	4.324 &	0.668 & 0.467 &	0.330 \\
    FoldFlow1 (base)  &  0.529 & 7.108 & 0.579 & 0.430 & 0.950 & 0.415 & 11.743 & 0.525 & 0.357 & 1.00 \\
    FoldFlow1 (sfm)  & 0.619 & 5.270 & 0.586 &	0.433 &	0.980 & 0.398 &	11.135 & 0.534 & 0.372 & 0.99 \\
    FoldFlow1 (ot)  &  0.528 &	6.877 &	0.582 &	0.392 &	0.900 & 0.418 &	10.78 &	0.559 &	0.365 &	0.99 \\
    Genie      &  0.367 & 13.699 &	0.431 &	0.264 &	1.00 & 0.251 &	24.453 &	0.238 &	0.229 &	1.00 \\

    % \midrule

    % MSA-1b & 0.921{\footnotesize$\pm$0.001} & 0.857$\pm$0.004 & 0.689$\pm$0.014 & 0.887$\pm$0.009 & 0.679$\pm$0.019 & 0.557$\pm$0.025 & 0.416 $\pm$ 0.050\\
    
    % \midrule

    \bottomrule
    \end{tabular}%
    }%
  \label{tab:backbone_design_sup}%
\end{table}%

In this section, we provide more detailed evaluation results of protein backbone design across additional length 200 and 400. The results is shown in Table.\ref{tab:backbone_design_sup}. In general, the performance of 200 and 400 backbones show a similar trend like other lengths.

\subsection{Antibody Design}
\label{appendix:antibody_design}
In this section, we will provide a detailed introduction to the evaluation of antibody design methods, including the overall evaluation concept, the variant to dyMEAN, the datasets used for training and testing, and the implementation for all evaluation metrics.

\subsubsection{Evaluation Concept}

As mentioned in the main text, antibody design can ultimately be simplified to the design of CDR-H3. Therefore, in this study, we evaluate the performance of different antibody design methods by evaluating the CDR-H3 sequences generated by these methods. Given the primary objective of this study is to assess the relative performance of various design models rather than the in vivo/vitro functionality of the antibodies they generate, we opted to directly evaluate the designed antibodies using their predicted structures. This approach is grounded in several considerations: firstly, it ensures a clear focus on evaluating the design methodology itself, independent of experimental constraints. Secondly, the significant time and resources required for extensive experimental validations, as well as the limitations of methods that can accurately simulate the real binding structure of antibodies, render in vivo/vitro assessments impractical. Direct evaluation of the designed structures presents a feasible and efficient strategy that aligns with the study’s goals and resource constraints while still providing valuable theoretical benchmarks for subsequent experimental investigations.

For methods capable of generating multiple antibodies for the same antigen, we generated 64 CDR-H3 sequences per antigen using each method and calculated the average performance across these different generated samples. Additionally, we also calculated the standard deviation of the performance among different samples generated for a single antigen.

\subsubsection{Variant of dyMEAN}
\label{appendix:antibody_design_dyMEANFixFR}
Unlike other methods, which are designed to accept the true structure of the antibody-antigen complex and generate the missing CDR-H3 region, dyMEAN is set up to accept only the structure of the antigen and the sequence of the non-CDR-H3 regions of the antibody. Therefore, the model needs to both generate the CDR-H3 region and predict the overall structure of the antibody as well as the binding pose between the antibody and antigen. Incorrect pose estimation can severely affect the interactions between CDR-H3 and the antigen, making a direct comparison between dyMEAN and other methods unfair. To compare dyMEAN with other methods more fairly, we made some modifications to dyMEAN by providing the true structure of the non-CDR-H3 regions of the antibody and the binding pose, aligning dyMEAN with the other methods. In dyMEAN-FixFR, we also used Rosetta \citep{alford2017rosetta} to repack the side chains, consistent with other methods, to avoid the influence of the side chains generated by dyMEAN on the evaluation results. Additionally, we introduced some randomness in the initialization of the structure, which allows dyMEAN-FixFR to generate multiple different antibodies for the same antigen.

\subsubsection{Dataset}
\label{appendix:antibody_design_dataset}
To retrain all the methods for a fair comparison, we use antibody-antigen complex structural data from the SAbDab dataset under the IMGT scheme \citep{lefranc2009imgt} as the training dataset. We collected antigen-antibody complexes with both heavy and light chains and protein antigens. We then discarded duplicate data with the same CDR-L3 and CDR-H3 sequence. The remaining complexes are used to cluster via MMseqs2 \citep{steinegger2017mmseqs2} with 40\% sequence similarity as the threshold based on the CDR-H3 sequence of each complex. Finally, we select the clusters that do not contain complexes in the RAbD dataset and split the complexes into training and validation sets with a ratio of 9:1 (1786 and 193 complexes respectively). 

The test set contains 55 antibody-antigen complexes extracted from the RAbD dataset. The original RAbD dataset contains 60 antibody-antigen complexes. In this study, we hope that the evaluation of antibody design methods is based on antibodies that contain both light and heavy chains, and simultaneously the antigen contains at least one protein chain. In practice, \textbf{2ghw} and \textbf{3uzq} lack light chains, while \textbf{3h3b} lack heavy chains. \textbf{5d96} is excluded because of the incorrect chain ID information in rabd\_summary.jsonl\footnote{\url{https://github.com/THUNLP-MT/MEAN/blob/main/summaries/rabd\_summary.jsonl}}, where heavy chain \textit{J} and light chain \textit{I} do not bind to antigen chain \textit{A}. \textbf{4etq} is excluded as HERN reported an error when running for this complex.

\subsubsection{Implementation of Evaluation Metrics}
\label{appendix:antibody_design_metrics}

\textbf{[Accuracy]}:
\begin{itemize}[leftmargin=*]
\item AAR: For the calculation of AAR (Amino Acid Recovery Rate), similar to existing work, we calculated the number of residues in the generated CDR-H3 sequences that match the natural antibody.
\item RMSD: In the calculation of RMSD (Root Mean Square Deviation), we measured the RMSD of the generated and natural antibodies in the CA coordinates of the CDR-H3 region. For methods other than dyMEAN, since their task setting provides the true binding pose of the antibody FR region and antigen, there is no need to align the generated structure with the natural structure when calculating RMSD. For dyMEAN, we aligned the 2 FR residues at each end of CDR-H3 with the corresponding residues in the natural antibody, applied the obtained transformation to CDR-H3, and then calculated the RMSD.
\item TM-score: We calculated the TM-score only for the CDR-H3 region. To this end, we saved the generated CDR-H3 part as a .pdb file and used \texttt{TMalign} 
 \citep{zhang2005tm} to calculate the TM-score between the generated CDR-H3 and the natural CDR-H3.
\end{itemize}

\textbf{[Functionality]}:
\begin{itemize}[leftmargin=*]
\item Binding Energy: The calculation of binding energy requires the all-atom structure of the protein, while most methods only generate the backbone atom structure. Therefore, we first used Rosetta to pack the missing side-chain atoms. Subsequently, we optimized the side-chains in the CDR-H3 region using Rosetta minimization while keeping the backbone structure unchanged to ensure that the CDR-H3 generated by the model reaches the minimum energy state in the binding environment with the antigen. During minimization, we set the step to 100 (we tried using more steps and repeats, although the energy did further decrease, the reduction was very limited and much smaller than the energy difference between different methods; however, the time consumption significantly increased). After minimization, we calculated the energy on the all-atom structure. Finally, we used the \texttt{InterfaceAnalyzer} in Rosetta to calculate the binding energy between CDR-H3 and the antigen.
\end{itemize}

\textbf{[Specificity]}:
\begin{itemize}[leftmargin=*]
\item SeqSim: SeqSim is defined as the average similarity between any sequence pairs among the generated sequences. First, we introduce the definition and implementation of similarity. The similarity between two sequences is defined as the percentage of matched amino acids over the aligned length after alignment (thus, this metric is affected by the length gap between the two sequences). Given that our goal is to calculate the number of matches rather than the matching score and that the two ends of CDR-H3 are fixed to FR3 and FR4, we need an alignment method that: (1) assigns a score of 1 for matches, and 0 for gaps and mismatches; (2) does not introduce gaps at the two ends of CDR-H3. We used the \texttt{PairwiseAligner} in Biopython \citep{cock2009biopython} for sequence alignment, setting \texttt{match\_score} to 1, all other scores to 0, and the \texttt{end\_gap\_score} to \textit{-inf} so that the alignment process meets our requirements. For methods that generate only one antibody per antigen, we directly calculate the average SeqSim among the 55 generated CDR-H3 sequences as SeqSim-outer. For methods that generate multiple antibodies, we calculate the average SeqSim between two sets of sequences generated for two antigens as SeqSim-outer and also calculate the average SeqSim within each set as SeqSim-inner.
The formulas for calculating SeqSim-outer and SeqSim-inner are as follows:
\begin{align}
\text{SeqSim-outer}&=\frac{1}{N*(N-1)*M^2}\sum_{i=1}^{N}\sum_{j=1|j\ne i}^{N}\sum_{x=1}^{M}\sum_{y=1}^{M}\text{SeqSim}(s_i^x, s_j^y),\label{eq:SeqSim_outer}\\
%\end{align}
%\begin{align}
\text{SeqSim-inner}&=\frac{1}{N*M*(M-1)}\sum_{i=1}^{N}\sum_{x=1}^{M}\sum_{y=1|y\ne~x}^{M}\text{SeqSim}(s_i^x, s_i^y),\label{eq:SeqSim_inner}
\end{align}
where $N$ denotes the number of antigens in the test set ($N$=55 in this study), $M$ denotes the number of samples generated for each antigen ($M$=64 in this study), and $s_i^x$ represents the $x$-th CDR-H3 sequence generated for the $i$-th antigen.
\item PHR: PHR is the proportion of hydrophobic residues in the generated CDR-H3 sequences. Although both PHR and SeqSim are used to represent the specificity of antibody design methods, they focus on different aspects. Thus, the same method may exhibit different tendencies in these two metrics (SeqSim can be understood as an evaluation of the method's specificity, while PHR is an evaluation of the specificity of the generated antibodies. When SeqSim performs poorly, the performance of PHR is of limited significance). For example, AbDPO achieves high SeqSim-outer but does not perform well in PHR. This indicates that AbDPO can specifically design antibodies for different antigens, but these antibodies contain many hydrophobic residues, leading to potential nonspecific interactions with multiple proteins.
\end{itemize}

\textbf{[Rationality]}:
\begin{itemize}[leftmargin=*]
\item CN-Score: To evaluate the consistency of the peptide bond length of generated antibodies with that of natural antibodies, we fit a Kernel Density Estimation (KDE) function using the length of peptide bonds found within the CDR-H3 regions of natural antibodies. The density of the generated peptide bond length, CN-Score, is used to represent the consistency. For generated peptide bonds shorter than the minimum natural peptide bond length or longer than the maximum, the density is defined as 0. The final CN-Score for a generated antibody is defined as the average density of the lengths of all its peptide bonds. It is important to note that the length variation of peptide bonds is very small, which leads to a very narrow distribution of natural peptide bond lengths. When the generated peptide bond length deviates slightly from the average length (1.3310), its density in the KDE function will sharply decrease, which explains why all methods show a significant difference in CN-Score compared to natural antibodies.
\item Clashes: Although atomic clashes within proteins mainly occur between the side chains, most methods do not generate the side chains of residues. Using packing methods to complete side chains can always find a side-chain conformation with the fewest clashes through extensive searching. Therefore, we instead evaluate the potential clash level in the generated structures rather than the specific number of clashes. To do this, we calculate the CA distance between two residues; when the CA-CA distance between two residues not connected by a covalent bond is less than the minimum CA-CA distance commonly found in covalently bonded residues (3.6574, derived from the CA-CA distance statistics in the CDR-H3 region of the RAbD dataset), we consider these two residues to have potential clashes. We then calculate the number of residue pairs with distances below this threshold to measure the clash level in the generated structures. The difference between Clashes-inner and Clashes-outer is: Clashes-inner measures the clash level within the generated CDR-H3 structure, while Clashes-outer measures the clash level between the generated CDR-H3 structure and other components, including the antigen, the heavy chain FR region, and the light chain of the antibody.
\item SeqNat: To measure how close the designed CDR-H3 sequences are to natural sequences, we used the pLL predicted by the AntiBERTy model. We input the entire heavy chain sequence into the model, which means that AntiBERTy makes prediction based on the entire heavy chain of the antibody, but unlike the standard procedure in AntiBERTy, the pLL calculation area is only within the CDR-H3 region (the standard procedure calculates pLL over the entire input sequence).
\item Total Energy: Before calculating the total energy, we performed the same energy optimization process on the designed CDR-H3 regions as described in the Functionality section. We then used Rosetta's full atom score function with the default weights from REF15 \citep{alford2017rosetta} to calculate the total energy of each residue in the CDR-H3 region. The Total Energy of the CDR-H3 region is defined as the sum of the total energy of all its residues.
\item scRMSD: In this metric, we used IgFold to predict the structure of the generated sequences. IgFold predicts the structure based on the sequence pair of the antibody's light and heavy chains (although the region we evaluate only exists in the heavy chain, and IgFold also supports single-chain input, we found that inputting two chains results in higher accuracy). The real structure of the non-CDR-H3 regions of the antibody was also provided as a template to obtain the initial predicted structure. We then used the Kabsch algorithm to align the non-CDR-H3 regions of the heavy chain with the real structure and applied the resulting transformation to the predicted CDR-H3 structure. This aligns the predicted CDR-H3 structure to its original complex. At this point, the CA-RMSD between the predicted CDR-H3 structure and the real structure in the RAbD dataset is 1.95. The structure predicted by IgFold is unrelated to the antigen, and since the antibody undergoes conformational changes in the binding interface after binding with the antigen, we used Rosetta to relax the predicted CDR-H3 in the presence of the antigen. The relaxation involves changes in both the backbone and side-chain structures. Specifically, we repeated relaxation runs five times for each structure predicted by IgFold, with 200 steps each time, and selected the structure with the lowest energy as the final predicted structure. At this stage, the CA-RMSD with the real structure decreased to 1.77. We then calculated the RMSD of the CA coordinates between the predicted structure and the backbone CA coordinates generated by the model, which is referred to as scRMSD.
\end{itemize}

\subsection{Protein Conformation Prediction}

In this section, we provide further details on the datasets, evaluation metrics, and model implementations used in the benchmark for \textit{Protein Conformation Prediction}.

\subsubsection{Datasets}

\begin{itemize}[leftmargin=*]
    \item \textbf{CAMEO2022} consists of 183 single protein chains collected from CAMEO targets between August and October 2022, with sequence lengths of less than 750 amino acids, following \cite{jingEigenFoldGenerativeProtein2023}. Protein sequences and structures were extracted from the mmCIF files available at the RCSB Protein Data Bank (https://www.rcsb.org/, \cite{berman2000rcsb}) using customized scripts. One of the proteins (PDB ID: 8AHP, chain A) has since been superseded by a new PDB entry 8QCW and we have replaced this chain with the updated record.
    \item \textbf{Apo-holo} consists of 91 single chain proteins curated by \cite{saldano2022impactapo}. The protein sequences and the structures of both \textit{apo} and \textit{holo} conformations were extracted using the same pipeline as in CAMEO2022. Follwoing \cite{jingEigenFoldGenerativeProtein2023}, we use the sequences of the \textit{apo} structures as the primary sequence for model inference.
    \item \textbf{BPTI} is a 58 amino acids protein whose dynamics have been extensively studied through long-time MD simulations \citep{shaw2010atomic}. We use the structures of the cluster centers identified in the MD study as the reference structures for evaluation.
    \item \textbf{ATLAS} is a recently published dataset containing triplicated 100 ns MD simulations for 1,390 diverse single-chain proteins \citep{yannATLASProteinFlexibility2024}. In this work, we use a subset of 82 proteins whose PDB entries were deposited after May 1, 2019, following \cite{jingAlphaFoldMeetsFlow2023}. ``Protein-only'' trajectories were downloaded from the ATLAS database \footnote{\url{https://www.dsimb.inserm.fr/ATLAS/index.html}} for evaluation.
\end{itemize}

\subsubsection{Evaluation Metrics}
\label{apdx:conf_metrics}

\textbf{[Accuracy]}

We evaluate the accuracy by comparing the generated conformations with reference structures. Specifically, TMscore, RMSD, GDT-TS are calculated using \texttt{TMscore} \cite{zhang2004TMscore} obtained from \url{https://zhanggroup.org/TM-score/}. We use \texttt{-seq} option to align sequences before structural alignment. lDDT scores are calculated using the original implementation \cite{mariani2013lddt}, available from \url{https://swissmodel.expasy.org/lddt/downloads/}.

In multiple-state prediction, we evaluate the accuracy of predicting a specific state by the best accuracy among the generated samples. For example, \textit{apo}-TM) is the highest TM-score between samples (N=20 in the benchmark) and the reference \textit{apo} structure. Following \cite{jingEigenFoldGenerativeProtein2023}, we evaluate the overall ability to predict multiple states using ``ensemble accuracy'', which is the average of accuracy for each state, measured by TM-score or RMSD.

For the accuracy metrics in distribution prediction task (flexibility, distributional accuracy, ensemble observables) , we follow the implementation \footnote{\url{https://github.com/bjing2016/alphaflow/blob/master/scripts/analyze\_ensembles.py}} of \cite{jingAlphaFoldMeetsFlow2023} with a modification to explicitly align atoms in the \texttt{mdtraj}'s \texttt{topology} objects between sample and reference structures. Below is an overview of these metrics: \textbf{Flexibility} (or Predicting flexibility in \cite{jingAlphaFoldMeetsFlow2023}) reflects how accurately the model predicts the diversity of proteins (Pairwise RMSD) or atoms (RMSF). This is measured by the Pearson correlation $r$ between the diversity measure of the model-generated samples (\textit{sample distribution} for short) and the reference MD samples (\textit{reference distribution} for short). \textbf{Distributional accuracy} directly compare the similarity between the sample and reference distributions. 
 RMWD is the root mean Wasserstein distance between the distributions of aligned coordinates, modeled as multivariate Gaussians. In this benchmark, we report only the total distance without translation and variance decomposition. Additionally, we evaluate the Wasserstein-2 distance of the conformational distribution in the first two PCA dimensions (of aligned coordinates), as well as the cosine similarity between the PCA components of the sample and reference distributions. Lastly, \textbf{ensemble observables} include metrics comparing specific observables (i.e., properties of interest) in the sample and reference distributions, particularlly, residue-residue contacts (weak or transient) and residue exposures (e.g., surface residue that contacts the solvent). Jaccard similarity and Mutual Information (MI) are used to compare these observables.

\textbf{[Diversity]}

Diversity is evaluated by the average pairwise structural similarity among the generated samples for a protein, measured using TM-score or RMSD. To reduce computation time, we randomly sample 100 pairs of structures for this calculation.

\textbf{[Quality]}

The quality of generated conformation structures are assessed by three backbone structural violations: CA-clash \%, CA-break \%, and PepBond-break \%.

\begin{itemize}[leftmargin=*]
    \item CA-clash \% is the rate of clashes between alpha-carbon atoms. A \textit{clash} is determined if the distance between a pair of alpha carbon atoms is less than 3.0 Å, similar to \cite{luStr2StrScorebasedFramework2024}. And CA-clash \% is calculated as
    $$
    \text{CA-clash \%} = \frac{\text{number of residues with clashes}}{\text{sequence length}} \times 100.
    $$
    \item CA-break \% is the rate of two connecting residues are too distant, leading to potential bond break. We determine a \textit{break} if the distance between two connecting residues is greater than 4.2~Å and CA-break \% is calculated as 
    $$
    \text{CA-break \%} = \frac{\text{number of connecting CA-CA pairs with break}}{\text{sequence length} - 1} \times 100.
    $$
    \item PepBond-break \% specifically evaluates the potential peptide bond (C-N) break between connecting residues, providing a more rigorous metric about inter-residue disconnection than CA-break \%. We use a maximum peptide length threshold of 1.4~Å to determine a chain break, as in the Biopython package\footnote{\url{https://biopython.org/docs/dev/api/Bio.PDB.internal\_coords.html\#Bio.PDB.internal\_coords.IC\_Chain}}. Similarily, PepBond-break \%  is calculated as 
    $$
    \text{PepBond-break \% } = \frac{\text{number of C-N bond break}}{\text{sequence length} - 1} \times 100.
    $$
\end{itemize}

\subsubsection{Model implementations}

\begin{itemize}[leftmargin=*]
    \item AlphaFold2 \citep{Jumper2021}: We used the ColabFold implementation \cite{mirditaColabFoldMakingProtein2022} for AlphaFold2 inference, with input MSAs obtained using the Colab pipeline. All five models (with pTM) were used to predict five candidate structures, and the structure with the highest pLDDT confidence score was selected for performance evaluation. All models were run with default settings, and no templates were provided.
    \item OpenFold \citep{Ahdritz2022.11.20.517210openfold}: We used \texttt{openfold v2.0.0} for inference with their pretrained OpenFold weights (with pTM). The same MSAs as those used for AlphaFold2 were provided as inputs. Since only one checkpoint (\texttt{finetuning\_no\_templ\_ptm\_1}) corresponding to the model configuration \texttt{model\_3\_ptm} is available, we generated three structures using three random seeds and made a total of 5 predictions. The structure with the highest pLDDT score was selected for performance evaluation. Templates were not provided for inference.
    \item ESMFold \citep{lin2023}: We use the public ESM repository for inference with the model \texttt{esm.pretrained.esmfold\_v1()}. Since EMSFold predictions are deterministic, we generated only one structure per protein for performance evaluation.
    \item RoseTTAFold2 \citep{Baek2023rosettafold2}: We follow their official repository and instructions for inference, with the same MSA as AlphaFold2 and OpenFold. No templates were provided. Only one structure per protein was predicted for performance evaluation. 
    \item EigenFold \citep{jingEigenFoldGenerativeProtein2023}: We follow the official repository, weights, and the setups provided by the authors for inference. In the \textit{protein folding} task, we sampled 5 structures for each protein and selected the one with the highest ELBO estimation for performance evaluation. Because EigenFold can not predict sequences containing unknown amino acids (labeled 'X'), we removed the 'X' in the input sequences, as done in the original implementation. This adjustment might introduce some performance difference due to inference with slightly different sequences.
    \item MSA-subsampling \citep{delalamoSamplingAlternativeConformational2022}: We implemented MSA-subsampling using the \texttt{openfold v2.0.0} package by adjusting the two configuration parameters, \texttt{max\_msa\_clusters} and \texttt{max\_extra\_msa}, following \cite{delalamoSamplingAlternativeConformational2022}. Specifically, we refer to \texttt{max\_extra\_msa} as the MSA depth and set \texttt{max\_msa\_clusters} to half that depth, while keeping other OpenFold settings at their default values. The original MSAs were obtained using the same ColabFold pipeline as in AlphaFold2.
    \item Str2Str \citep{luStr2StrScorebasedFramework2024}: We followed the official implementation of Str2Str and used OpenFold-predicted structure as the initial structures. Ensemble results were collected by uniformly sampling from $t$ values. For BPTI, we used the author-recommended noising schedule with maximum forward time of $T_\text{max} = 0.15 (t=0.10, 0.15)$. For \textit{apo}-\textit{holo} and ATLAS datasets, we experimented with $T_\text{max}=0.1 (t=0.06, 0.08, 0.10, 0.12, 0.14)$ and  $T_\text{max}=0.3 (t=0.06, 0.12, 0.18, 0.24, 0.30)$ for both the SDE and ODE models. 
    \item AlphaFlow/ESMFlow \citep{jingAlphaFoldMeetsFlow2023}: We used the official repository and released model weights for inference. The MSAs for AlphaFlow models were obtained through ColabFold's pipeline.
    \item ConfDiff \citep{wang2024proteinconfdiff}: We followed the authors' implementation and used the released weights for inference. In this benchmark, we used recycle3 representations for both ConfDiff-Open and ConfDiff-ESM models. The energy and force guidance models are dataset-specific and are only available for the BPTI dataset with ESMFold representations.
\end{itemize}

\subsubsection{Extended discussion on protein conformation models}
\label{apdx:model_compare}

Recent works have explored various strategies to extend current structure prediction models to generate diverse and plausible conformations. Below, we briefly discuss the key differences among the studies and strategies compared in this benchmark:

\begin{itemize}[leftmargin=*]
    \item \textbf{Perturbing the input of folding models.} While AlphaFold2 is designed to predict the folded structure of a protein, several studies have proposed pertubing its MSA input to generate alternative structures (as conformations) without re-training the model \citep{delalamoSamplingAlternativeConformational2022,wayment2024predictingMSA}. In this benchmark, we assess MSA subsampling, a method that reduces the number of input MSAs (referred to as ``depth'') by subsampling the full MSA, enabling the prediction of different "folded" structures due to the depletion of the input information.
    
    \item \textbf{Perturbing folded structures.} Instead of perturbing the input to a folding model, Str2Str~\citep{luStr2StrScorebasedFramework2024} perturbs the structure predicted from a folding model. It uses a structure-only diffusion model (i.e., a backbone design model) to perturb the input structure through a forward-backward diffusion process. The level of perturbation is controlled by the maximum diffusion time, $T_\text{max}$. They also used ensembling by sampling at various diffusion times $t \le T_\text{max}$.
    
    \item \textbf{Training generative models on large-scale structural data from experiments or simulations.} A more direct approach involves training sequence-conditioned generative models using diffusion or flow frameworks. EigenFold~\citep{jingEigenFoldGenerativeProtein2023}, AlphaFlow~\citep{jingAlphaFoldMeetsFlow2023}, and ConfDiff~\citep{wang2024proteinconfdiff} follow similar approaches by fine-tuning a diffusion time $t$-dependent score or denoising model based on folding models, using structural data from PDB. Specifically, AlphaFlow finetunes all layers of AlphaFold2, while EigenFold and ConfDiff use pretrained representations from folding models and train a lightweight add-on module for score or denoising prediction. Despite adopting a generative framework, models solely trained on PDB data are limited in predicting conformational distributions. To address this, AlphaFlow and ConfDiff further fine-tuned their models on a recent MD dataset containing densely sampled conformations for proteins (see Atlas in the Datasets section). 
    % Distributed Graphormer~\citep{zheng2024predictingDiG} adopted a similar approach, incorperating pre-training followed by fine-tuning on MD data. 

    \item \textbf{Integrate physical priors in conformational training or sampling.} 
    Due to limited availability of large-scale protein conformation data from MD simulation, some models have explored integrating structural and physical priors during training. ConfDiff \citep{wang2024proteinconfdiff} introduced two guidance techniques to improve conformational sampling: (1) classifier-free guidance, which combines a sequence-conditioned conformation model with an unconditional (structure-only) model to explore conformational space (ConfDiff-ClsFree), and (2) energy/force guidance, which directs sampling toward regions with lower potential energy (ConfDiff-Energy/Force) through auxiliary prediction modules for intermediate energy/force guidance. However, such physical prediction modules are dataset-specific and requires training additional modules. 
    % Distributed Graphormer~\citep{zheng2024predictingDiG} proposed a physics-informed diffusion training loss to augment the diffusion training with forces evaluated by a force field model.

\end{itemize}

\subsubsection{Additional evaluation results}
\label{apdx:full_eval}
Here we include the complete evaluation results for multiple-state prediction (BPTI and apo-holo) and distribution prediction (Atlas).
% \quad

\begin{table}[h]\footnotesize
  \centering\setlength{\tabcolsep}{2pt}
  \caption{Complete performance on the multiple-state prediction of BPTI. Accuracy metrics (RMSDens, RMSD Cluster 3) are reported as the mean and standard deviations from 20 bootstrap samples with replacement, at different sample sizes ($N=10\sim1000$). Diversity and Quality scores are evaluated based on 1,000 conformations for each model. The \textbf{best} performance is highlighted in bold, and the \uline{second-best} is underlined. ``N/A'' indicates not applicable due to model resolution. RMSD is measured in Å.}

\resizebox{\textwidth}{!}{%

\begin{tabular}{lcccccccccc}
\toprule
 & \multicolumn{3}{c}{\textbf{RMSDens} ↓} & \multicolumn{3}{c}{\textbf{RMSD Cluster 3} ↓} & \textbf{Diversity} & \multicolumn{2}{c}{\textbf{Quality}} \\
 \cmidrule(lr){2-4}  \cmidrule(lr){5-7} \cmidrule(lr){8-8}  \cmidrule(lr){9-10}
 & \textbf{N=10} & \textbf{N=100} & \textbf{N=1000} & \textbf{N=10} & \textbf{N=100} & \textbf{N=1000} & \textbf{\specialcell{Pairwise\\RMSD}} & \specialcell{\textbf{CA}\\\textbf{clash\%} ↓} & \specialcell{\textbf{PepBond}\\\textbf{break\%}↓} \\
\midrule

EigenFold & \uline{1.56$\pm$0.02} & 1.50$\pm$0.01 & 1.46$\pm$0.00 & 2.54$\pm$0.03 & 2.48$\pm$0.01 & 2.46$\pm$0.01 & 0.85 & 1.4 & N/A \\
MSA-depth256 & 1.58$\pm$0.01 & 1.54$\pm$0.01 & 1.52$\pm$0.01 & 2.51$\pm$0.02 & 2.48$\pm$0.01 & 2.44$\pm$0.01 & 0.20 & \textbf{0.0} & 9.2 \\
MSA-depth64 & 1.60$\pm$0.01 & 1.55$\pm$0.02 & 1.51$\pm$0.01 & 2.46$\pm$0.03 & 2.41$\pm$0.04 & 2.34$\pm$0.03 & 0.55 & \textbf{0.0} & 7.9 \\
MSA-depth32 & 1.66$\pm$0.03 & 1.54$\pm$0.04 & 1.41$\pm$0.02 & \textbf{2.43$\pm$0.06} & \textbf{2.19$\pm$0.16} & \textbf{1.85$\pm$0.05} & 2.14 & 0.6 & 10.6 \\
Str2Str-ODE ($T_\text{max}=0.15$) & 2.40$\pm$0.12 & 2.20$\pm$0.05 & 2.09$\pm$0.01 & 3.00$\pm$0.20 & 2.73$\pm$0.12 & 2.58$\pm$0.05 & 1.86 & \textbf{0.0} & 13.9 \\
Str2Str-SDE ($T_\text{max}=0.15$) & 2.76$\pm$0.16 & 2.46$\pm$0.08 & 2.26$\pm$0.04 & 3.26$\pm$0.25 & 2.86$\pm$0.25 & 2.55$\pm$0.16 & 3.60 & 0.3 & 16.0 \\
AlphaFlow-PDB & \textbf{1.53$\pm$0.03} & 1.46$\pm$0.01 & 1.41$\pm$0.01 & 2.48$\pm$0.04 & 2.43$\pm$0.01 & 2.40$\pm$0.01 & 0.86 & \textbf{0.0} & 13.2 \\
AlphaFlow-MD & 1.71$\pm$0.08 & 1.51$\pm$0.03 & 1.43$\pm$0.01 & 2.46$\pm$0.09 & 2.32$\pm$0.06 & 2.25$\pm$0.01 & 1.26 & \textbf{0.0} & 26.2 \\
ESMFlow-PDB & 1.59$\pm$0.04 & 1.49$\pm$0.02 & 1.42$\pm$0.01 & 2.49$\pm$0.03 & 2.41$\pm$0.03 & 2.34$\pm$0.01 & 0.74 & \textbf{0.0} & 6.0 \\
ESMFlow-MD & 1.68$\pm$0.06 & 1.47$\pm$0.04 & 1.39$\pm$0.03 & \uline{2.44$\pm$0.11} & \uline{2.27$\pm$0.10} & 2.18$\pm$0.02 & 1.17 & \textbf{0.0} & 14.3 \\
ConfDiff-Open-MD & 1.64$\pm$0.05 & 1.50$\pm$0.02 & 1.43$\pm$0.02 & 2.50$\pm$0.05 & 2.38$\pm$0.04 & 2.31$\pm$0.02 & 1.37 & 0.2 & \textbf{4.6} \\
ConfDiff-Open-ClsFree & 1.66$\pm$0.06 & 1.50$\pm$0.04 & \uline{1.37$\pm$0.02} & 2.56$\pm$0.07 & 2.39$\pm$0.17 & \uline{2.02$\pm$0.10} & 1.77 & 0.5 & 5.5 \\
ConfDiff-ESM-MD & 1.62$\pm$0.04 & 1.47$\pm$0.02 & 1.40$\pm$0.01 & 2.45$\pm$0.09 & 2.32$\pm$0.05 & 2.25$\pm$0.02 & 1.42 & 0.1 & \uline{5.0} \\
ConfDiff-ESM-ClsFree & 1.57$\pm$0.04 & \uline{1.45$\pm$0.02} & 1.40$\pm$0.01 & 2.48$\pm$0.04 & 2.40$\pm$0.03 & 2.34$\pm$0.02 & 1.80 & 0.5 & 7.5 \\
ConfDiff-ESM-Energy & 1.61$\pm$0.03 & 1.46$\pm$0.02 & 1.42$\pm$0.01 & 2.51$\pm$0.05 & 2.44$\pm$0.03 & 2.40$\pm$0.01 & 1.22 & 0.1 & 7.5 \\
ConfDiff-ESM-Force & 1.58$\pm$0.04 & \textbf{1.43$\pm$0.03} & \textbf{1.36$\pm$0.01} & \uline{2.44$\pm$0.06} & 2.35$\pm$0.05 & 2.24$\pm$0.06 & 1.76 & 0.1 & 8.9 \\
\bottomrule
\end{tabular}

}%

\label{tab:conf_bpti}%
\end{table}%
\newpage
\quad
\begin{table}[h]\footnotesize
  \centering\setlength{\tabcolsep}{2pt}
  \caption{Performance on the conformation prediction task for the \textit{apo-holo} dataset. \textit{apo}/\textit{holo}-TM represents the maximum TM-score of the samples relative to the reference \textit{apo}/\textit{holo} structure. Twenty conformations were sampled for each protein, and the results are reported as mean/median across 91 proteins. The \textbf{best} performance is highlighted in bold, and the \uline{second-best} is underlined. ``N/A'' indicates not applicable due to model resolution.}
\resizebox{0.8\textwidth}{!}{%

% No TMmin
\begin{tabular}{lcccccc}
\toprule
 & \multicolumn{3}{c}{\textbf{Accuracy}} & \textbf{Diversity} & \multicolumn{2}{c}{\textbf{Quality}} \\
  \cmidrule(lr){2-4}  \cmidrule(lr){5-5} \cmidrule(lr){6-7}
 & \textbf{\textit{apo}-TM} ↑ & \textbf{\textit{holo}-TM} ↑  & \textbf{TMens} ↑ & \textbf{Pairwise TM} & \textbf{CA clash \%} ↓ & \textbf{\specialcell{PepBond break \%}} ↓ \\
 
\midrule
\rowcolor{gray!15}
\textit{apo} model & 1.000 & 0.790 & 0.895 & N/A & N/A & N/A \\
EigenFold & 0.831 & 0.864 & 0.847 & 0.907 & 3.6 & N/A \\
MSA-depth256 & 0.845 & \uline{0.889} & \uline{0.867} & 0.978 & \uline{0.2} & 4.6 \\
MSA-depth64 & 0.844 & 0.883 & 0.863 & 0.950 & \uline{0.2} & 5.7 \\
MSA-depth32 & 0.824 & 0.857 & 0.841 & 0.864 & \uline{0.2} & 8.9 \\
Str2Str-ODE ($T_\text{max}=0.1$) & 0.762 & 0.778 & 0.770 & 0.954 & \uline{0.2} & 14.0 \\
Str2Str-ODE ($T_\text{max}=0.3$) & 0.766 & 0.781 & 0.774 & 0.872 & \uline{0.2} & 14.7 \\
Str2Str-SDE ($T_\text{max}=0.1$) & 0.682 & 0.693 & 0.688 & 0.760 & \uline{0.2} & 22.6 \\
Str2Str-SDE ($T_\text{max}=0.3$) & 0.680 & 0.689 & 0.684 & 0.639 & \uline{0.2} & 21.1 \\
AlphaFlow-PDB & \uline{0.855} & \textbf{0.891} & \textbf{0.873} & 0.924 & 0.3 & 6.6 \\
AlphaFlow-MD & \textbf{0.857} & 0.863 & 0.860 & 0.894 & \uline{0.2} & 20.8 \\
ESMFlow-PDB & 0.849 & 0.882 & 0.866 & 0.935 & 0.3 & 4.8 \\
ESMFlow-MD & 0.851 & 0.864 & 0.858 & 0.897 & \textbf{0.1} & 10.9 \\
ConfDiff-Open-PDB & 0.847 & 0.886 & \uline{0.867} & 0.909 & 0.5 & 5.5 \\
ConfDiff-Open-ClsFree & 0.838 & 0.879 & 0.859 & 0.870 & 0.8 & 5.8 \\
ConfDiff-Open-MD & 0.839 & 0.874 & 0.857 & 0.863 & 0.4 & 6.8 \\
ConfDiff-ESM-PDB & 0.845 & 0.873 & 0.859 & 0.890 & 0.5 & \textbf{4.1} \\
ConfDiff-ESM-ClsFree & 0.837 & 0.864 & 0.850 & 0.846 & 0.7 & 4.6 \\
ConfDiff-ESM-MD & 0.836 & 0.862 & 0.849 & 0.846 & 0.3 & \textbf{4.1} \\

\bottomrule
\end{tabular}

}%

\label{tab:conf_apo}%
\end{table}%
\newpage
\begin{table}[h]\footnotesize
  \centering\setlength{\tabcolsep}{2pt}
  \caption{Performance on distribution prediction for the ATLAS test set. A total of 250 conformations were sampled for each protein, and the median values across 82 proteins are reported. The \textbf{best} performance is highlighted in bold, and the \uline{second-best} is underlined. *These metrics require all-atom or backbone predictions; therefore, EigenFold and Str2Str do not have sufficient resolution for evaluation (indicated as ``N/A'').}

\resizebox{\textwidth}{!}{%
\begin{tabular}{lccccccccc}
\toprule
 & \multicolumn{2}{c}{\textbf{Diversity}} & \multicolumn{3}{c}{\textbf{Flexibility: \textit{Pearson} $r$ on}} & \multicolumn{4}{c}{\textbf{Distributional accuracy}} \\
 \cmidrule(lr){2-3}  \cmidrule(lr){4-6} \cmidrule(lr){7-10}
 & \textbf{\specialcell{Pairwise\\RMSD}} & \textbf{*RMSF} & \specialcell{\textbf{Pairwise}\\\textbf{RMSD} ↑ } & \specialcell{*\textbf{Global}\\\textbf{RMSF} ↑ }  & \specialcell{*\textbf{Per target}\\\textbf{RMSF} ↑ } & \specialcell{*\textbf{RMWD} ↓} & \specialcell{\textbf{MD PCA}\\\textbf{W2} ↓} & \specialcell{\textbf{Joint}\\\textbf{PCA W2} ↓} & \specialcell{\textbf{PC sim}\\\textbf{$>$ 0.5 \%}↑ } \\
\midrule
\rowcolor{gray!15}
MD iid & 2.76 & 1.63 & 0.96 & 0.97 & 0.99 & 0.67 & 0.73 & 0.71 & 93.9 \\
\rowcolor{gray!15}
MD 2.5ns & 1.54 & 0.98 & 0.89 & 0.85 & 0.85 & 2.22 & 1.55 & 1.89 & 36.6 \\
EigenFold & 5.96 & N/A & -0.03 & N/A & N/A & N/A & 2.31 & 7.96 & 12.2 \\
MSA-depth256 & 0.83 & 0.53 & 0.25 & 0.34 & 0.59 & 3.60 & 1.79 & 2.91 & 29.3 \\
MSA-depth64 & 2.03 & 1.51 & 0.25 & 0.30 & 0.57 & 4.00 & 1.94 & 3.34 & 18.3 \\
MSA-depth32 & 5.70 & 7.96 & 0.08 & 0.17 & 0.53 & 6.09 & 2.56 & 5.70 & 17.1 \\
Str2Str-ODE ($T_\text{max}=0.1$) & 1.66 & N/A & 0.13 & N/A & N/A & N/A & 2.14 & 4.39 & 6.1 \\
Str2Str-ODE ($T_\text{max}=0.3$) & 3.15 & N/A & 0.13 & N/A & N/A & N/A & 2.19 & 4.80 & 9.8 \\
Str2Str-SDE ($T_\text{max}=0.1$) & 4.74 & N/A & 0.11 & N/A & N/A & N/A & 2.54 & 8.82 & 9.8 \\
Str2Str-SDE ($T_\text{max}=0.3$) & 7.54 & N/A & 0.01 & N/A & N/A & N/A & 3.24 & 12.28 & 7.3 \\
AlphaFlow-PDB & 2.58 & 1.20 & 0.27 & 0.46 & 0.81 & 2.97 & 1.61 & 2.61 & 37.8 \\
AlphaFlow-MD & 2.87 & 1.63 & \uline{0.53} & \uline{0.66} & \textbf{0.85} & \textbf{2.64} & 1.55 & \uline{2.29} & \uline{39.0} \\
ESMFlow-PDB & 2.99 & 1.68 & 0.14 & 0.27 & 0.71 & 4.15 & 1.87 & 3.61 & 28.0 \\
ESMFlow-MD & 3.33 & 2.13 & 0.19 & 0.30 & 0.76 & 3.61 & 1.66 & 3.25 & 25.6 \\
ConfDiff-Open-ClsFree & 3.68 & 2.12 & 0.39 & 0.54 & 0.83 & 2.91 & \uline{1.54} & 2.46 & \textbf{46.3} \\
ConfDiff-Open-PDB & 2.89 & 1.43 & 0.38 & 0.51 & 0.82 & 2.96 & 1.59 & 2.46 & 34.1 \\
ConfDiff-Open-MD & 3.43 & 2.21 & \textbf{0.59} & \textbf{0.67} & \textbf{0.85} & \uline{2.75} & \textbf{1.41} & \textbf{2.27} & 35.4 \\
ConfDiff-ESM-ClsFree & 4.04 & 2.84 & 0.31 & 0.43 & 0.82 & 3.78 & 1.73 & 3.07 & 37.8 \\
ConfDiff-ESM-PDB & 3.42 & 2.06 & 0.29 & 0.40 & 0.80 & 3.62 & 1.68 & 3.13 & 34.1 \\
ConfDiff-ESM-MD & 3.90 & 2.79 & 0.35 & 0.48 & 0.82 & 3.62 & 1.73 & 3.00 & 37.8 \\
\bottomrule
 & \multicolumn{4}{c}{\textbf{Ensemble observables}} & \multicolumn{2}{c}{\textbf{Quality}}  \\
  \cmidrule(lr){2-5}  \cmidrule(lr){6-7}
 & \specialcell{\textbf{Weak}\\\textbf{contacts $J$} ↑} & \specialcell{\textbf{Transient}\\\textbf{contacts $J$}↑} & \specialcell{*\textbf{Exposed} \\\textbf{residue $J$} ↑} & \specialcell{*\textbf{Exposed MI} \\\textbf{matrix $\rho$} ↑} & \specialcell{\textbf{CA clash}\\\textbf{\%} ↓} & \specialcell{*\textbf{PepBond}\\\textbf{break \%} ↓} &  & &  \\
\midrule
\rowcolor{gray!15} 
MD iid & 0.90 & 0.80 & 0.93 & 0.56 & 0.0 & 3.4 \\
\rowcolor{gray!15} 
MD 2.5ns & 0.62 & 0.45 & 0.64 & 0.25 & 0.0 & 3.4 \\
EigenFold & 0.36 & 0.19 & N/A & N/A & 5.6 & N/A \\
MSA-depth256 & 0.30 & 0.29 & 0.36 & 0.06 & \textbf{0.0} & 5.5 \\
MSA-depth64 & 0.38 & 0.28 & 0.40 & 0.16 & \textbf{0.0} & 7.6 \\
MSA-depth32 & 0.40 & 0.24 & 0.40 & 0.19 & 0.1 & 11.2 \\
Str2Str-ODE ($T_\text{max}=0.1$) & 0.42 & 0.18 & N/A & N/A & \textbf{0.0} & 12.1 \\
Str2Str-ODE ($T_\text{max}=0.3$) & 0.42 & 0.17 & N/A & N/A & \textbf{0.0} & 13.2 \\
Str2Str-SDE ($T_\text{max}=0.1$) & 0.40 & 0.13 & N/A & N/A & 0.1 & 21.9 \\
Str2Str-SDE ($T_\text{max}=0.3$) & 0.36 & 0.13 & N/A & N/A & 0.2 & 20.2 \\
AlphaFlow-PDB & 0.45 & 0.36 & 0.50 & 0.25 & 0.1 & 6.7 \\
AlphaFlow-MD & \uline{0.62} & \textbf{0.41} & \textbf{0.69} & \textbf{0.35} & \textbf{0.0} & 22.2 \\
ESMFlow-PDB & 0.42 & 0.30 & 0.46 & 0.21 & 0.2 & 5.1 \\
ESMFlow-MD & 0.55 & 0.34 & 0.57 & 0.29 & 0.1 & 10.9 \\
ConfDiff-Open-ClsFree & 0.58 & 0.36 & 0.60 & 0.28 & 0.8 & 5.7 \\
ConfDiff-Open-PDB & 0.50 & 0.36 & 0.54 & 0.25 & 0.5 & 5.6 \\
ConfDiff-Open-MD & \textbf{0.63} & \uline{0.39} & \uline{0.65} & \uline{0.33} & 0.5 & 6.5 \\
ConfDiff-ESM-ClsFree & 0.57 & 0.34 & 0.59 & 0.23 & 0.9 & \uline{4.3} \\
ConfDiff-ESM-PDB & 0.50 & 0.33 & 0.50 & 0.24 & 0.5 & \textbf{4.0} \\
ConfDiff-ESM-MD & 0.61 & 0.36 & 0.61 & 0.31 & 0.4 & \uline{4.3} \\
\bottomrule
\end{tabular}
}%
\label{tab:conf_atlas}%
\end{table}%j

\newpage
\bibliographystyle{iclr2025_conference}
\bibliography{iclr2025_conference}

\end{document}